\documentclass[12pt,preprint]{aastex}

\usepackage{epsfig}

\newcommand{\indeg}[1]{{#1}^{\circ}}

\def\vv{{\mathbf{v}}}
\def\vr{{\mathbf{r}}}
\def\vs{{\mathbf{J}}}
\def\vh{{\mathbf{h}}}
\def\vb{{\mathbf{B}}}

\def\phat{{\hat\psi}}

\newcommand{\runm}{\texttt{BP-m}}
\newcommand{\runh}{\texttt{BP-h}}



\begin{document}

\title{MHD Simulation of a Disk Subjected to Lense-Thirring Precession}

\author{Kareem A. Sorathia\altaffilmark{1}, Julian H. Krolik\altaffilmark{1}}

\and

\author{John F. Hawley\altaffilmark{2}}

\altaffiltext{1}{Department of Physics and Astronomy, Johns Hopkins University, Baltimore, MD 21218, USA} 

\altaffiltext{2}{Department of Astronomy, University of Virginia, Charlottesville VA 22904, USA}

\begin{abstract}
When matter orbits around a central mass obliquely with respect to the mass's spin axis,
the Lense-Thirring effect causes it to precess at a rate declining sharply with radius.
Ever since the work of Bardeen \& Petterson (1975), it has been expected that when a fluid
fills an orbiting disk, the orbital angular momentum at small radii should then align with
the mass's spin.  Nearly all previous work has studied this alignment under the assumption
that a phenomenological ``viscosity" isotropically degrades fluid shears in accretion
disks, even though it is now understood that internal stress in flat disks is due to
anisotropic MHD turbulence.  In this paper we report a pair of matched simulations, one
in MHD and one in pure (non-viscous) HD in order to clarify the specific mechanisms of
alignment.
As in the previous work, we find that disk warps induce radial flows that mix angular
momentum of different orientation; however, we also show that the speeds of these
flows are generically transonic and are only very weakly influenced by internal stresses
other than pressure.   In particular, MHD turbulence does not act in a manner consistent
with an isotropic viscosity.
When MHD effects are present, the disk aligns, first at small radii and
then at large; alignment is only partial in the HD case.  We identify the specific angular
momentum transport mechanisms causing alignment and show how MHD effects permit
them to operate more efficiently.  Lastly, we relate
the speed at which an alignment front propagates outward (in the MHD case) to the
rate at which Lense-Thirring torques deliver angular momentum at smaller radii.



\end{abstract}


\section{Introduction}
\label{sec:int}
There are many reasons why accretion disks around rotating masses may not be aligned with the
spin axis of the central mass.  If the matter is supplied from a companion star, the orbital plane
of the binary may be oblique; if the matter is supplied from the interstellar medium, its
mean orbital plane may similarly be inclined; if the matter is supplied from a tidally-disrupted
star, the orbital plane should be entirely uncorrelated with the black hole spin.  Whatever
the origin of the misalignment, general relativity predicts that there is a torque exerted
on the orbiting material.  To lowest post-Newtonian order, the torque on an orbiting ring of
radius $r$ and angular momentum ${\bf L}$ is $2 (G/c^2)\vs \times {\bf L}/r^3$ when the
spin angular momentum of the central mass is ${\bf J}$; the result is precession about the
central mass's spin axis at a rate $\omega = 2G |\vs|/(r^3 c^2)$.  The precession
rate is most interesting, of course, when $r$ is not an extremely large number of gravitational
radii $r_g \equiv GM/c^2$, so the effect is normally associated with black holes, or perhaps
neutron stars.

Because the torque grows so rapidly with smaller $r$, it has been supposed ever since the
truly seminal paper of \cite{BP75} that the strong differential precession at small radii
will induce internal disk friction, causing the inner part of the disk to settle
into the equatorial plane of the central mass's rotation.  \cite{PP83} invented an angular
momentum-conserving formalism to encompass this picture, in which they pointed out that
local disk warps can be smoothed hydrodynamically because the warps create radial
pressure gradients by shifting neighboring rings vertically relative to one another.
Radial fluid motions are then induced, which can mix the differently-oriented angular
momenta of the adjacent rings.  The question that arises, however, is how to relate the
radial velocities to the radial pressure gradients.  \cite{PP83} proposed that when
the disk is very thin, the flow velocities would be limited by the same ``viscosity"
accounting for angular momentum transport in flat disks, but operating isotropically
on all shears.
Conversely, they argued, when the disk is relatively thick, this ``viscosity" would
damp bending waves.
\cite{Pringle92} constructed a simpler vehicle for analyzing this geometrically complicated
problem, heuristically separating the angular momentum transport into two pieces.  The
first of these was the radial transport of angular momentum due directly to the action
of the isotropic viscosity.  The second was a
lumped-parameter description of the evolution of local warps, in which they were supposed
to be smoothed diffusively.  Following \cite{PP83}, \cite{Pringle92} argued that
the diffusivity for local warps would scale inversely with the putative isotropic
viscosity. \cite{O99} developed a nonlinear theory linking the mutual scaling of the
two transport coefficients.  When the viscosity
is normalized to the local pressure via the dimensionless coefficient $\alpha$, \cite{O99}
confirmed the expected inverse scaling for small values of $\alpha$, but found a somewhat more
complicated relation for larger values, provided $\alpha < 1$.  It is therefore natural
to describe the magnitude of the diffusion coefficient in the same pressure-normalized
fashion, scaling it in terms of $\alpha_2$ \citep{LP07}.  \cite{NP00} performed SPH
simulations in which the numerical diffusion of the algorithm provided an effective
isotropic viscosity and found behavior more or less in keeping with these expectations,
but \cite{LP07} and \cite{Lodato10}, using an explicit isotropic viscosity, argued that $\alpha_2$
was limited to be $\lesssim 3$, and also found that a diffusive description did not well match
the evolution of their simulations when the warp was ``nonlinear" (see \S~\ref{sec:warp} for
the definition of ``nonlinear" in this context).
Perhaps more surprisingly, the recent SPH simulations of \cite{Nixon12} develop sharp breaks
in the disk profile when the degree of misalignment is large.

There is, however, a fundamental worry concerning this entire approach: the assumption that
an isotropic viscosity acts in accretion disks.  For more than fifteen years
\citep{mri91,hgb,brand95,shgb96,bh98}, it has been clear that the angular momentum
transport governing accretion is {\it not} due to any sort of viscosity, but rather to MHD
turbulence driven by the magneto-rotational instability.  These stresses, although
related in the mean to orbital shear, are far from isotropic \citep{shgb96,HGK11}, do not scale
linearly with the shear, and do not respond in any direct way to fluctuating shears
\citep{Pessah08}.  In fact, when radial motions are sheared vertically in a magnetized
orbiting plasma, they are unstable when, as is the case here, the vertical scale of the
shear is longer than the distance an Alfven wave travels in a dynamical time \citep{mri91}.
All these contrasts call into question whether, or under what circumstances, MHD-derived
stresses might either limit radial flows or damp bending waves.  To date, only one
numerical simulation has been used to investigate the
effects produced by Lense-Thirring torques on disks with internal MHD turbulence
\citep{Fragile07}, but
its interpretability was limited by the nearness of the disk to the innermost stable
circular orbit, the disk's relatively large scale height, and the difficulty of
adequately resolving the MHD turbulence.
Thus, the applicability of this central assumption to the theory is still unclear.

In addition to this concern, there is also another reason to revisit the dynamics of
the Bardeen-Petterson problem.  Despite all the effort devoted to its study, there is
still no clear understanding of angular momentum flows during the process of inner-disk
alignment.  If the angular momentum given the disk material by the Lense-Thirring torque
remained with the material initially receiving it, the matter would simply precess around the
central mass's spin axis while very gradually drifting inward (this was, in fact, the
way \cite{BP75} originally envisioned it).  On the other hand, if hydrodynamic effects
redistribute the angular momentum given the disk by the torques \citep{PP83}, where does
it go?  Could the MHD turbulence carry the unaligned angular momentum a substantial
distance?  Moreover, why should that redistribution lead to alignment?   After all, averaged
over many precession periods, the net integrated angular momentum due to the torque goes to zero.

To begin answering these questions, we have performed a new MHD simulation of a
disk evolving under the influence of Lense-Thirring torques.  In its definition,
we have made the strategic decision to forego a fully relativistic treatment so
that the computational effort can be focused on resolving the MHD turbulence in a
reasonably thin disk and running for a sufficiently long time, rather than on the dynamical
complications of general relativity.  It therefore assumes Newtonian dynamics except for a
single term expressing the gravito-magnetic torque to lowest post-Newtonian order.
We have also chosen parameters such that the precession rate in the middle of
the disk is slower than the orbital period, but rapid enough that the mid-point
of the disk (if unencumbered by hydrodynamics) would precess through a full
rotation over the course of the simulation. 
The Newtonian approximation is an advantage here, too, because
such a comparatively rapid precession rate is found in a genuine relativistic context
only in the region not far outside the ISCO, where the warp dynamics
would be obscured by a mass inflow rate comparable to the precession frequency.

\section{Simulations}
\label{sec:sims}

The simulation code we employ is a contemporary translation (in Fortran-95) of the
3D finite-difference MHD code {\it Zeus} \citep{zeus1,zeus2}.  The magnetic field
is updated using the ``method of characteristics constrained transport (MOCCT)" algorithm
to maintain zero divergence to machine accuracy \citep{HawleyStone95}. 
The {\it Zeus} code solves the standard equations of Newtonian fluid dynamics, but we
augment its momentum equation with a term of the form $\rho \vv \times \vh$ to represent the
gravitomagnetic force per unit mass, where $\rho$ is the mass density, $\vv$ is the
fluid velocity, and
\begin{equation}
\vh = \frac{2\vs}{r^3} - \frac{6(\vs \cdot \vr)\vr}{r^5}.
\end{equation}
Here $\vs$ represents the magnitude and direction of the spin vector of the central mass
and $r$ is spherical radius.

In this paper we report two simulations, one employing full 3D MHD, but the other
purely hydrodynamic so that we may identify the special properties due to MHD through
contrasting the two.  The initial condition for the MHD simulation is a hydrostatic torus
orbiting a point-mass in Newtonian gravity
(see \cite{Hawley00})
defined by the parameters $q=1.65$, $r_{in} = 7.5$, $r_{\rm M} = 10$, $\rho_{\rm M} = 100$,
and $\Gamma = 5/3$.  That is, in this initial state the orbital frequency $\Omega \propto R^{-q}$
for cylindrical radius $R$, and the disk extends from an inner radius $r_{\rm in}$ to an
outer radius $r_{\rm out} \simeq 14$.  Its pressure maximum is found at $r_{\rm M}$, where
the density is $\rho_{\rm M}$.  We assume an adiabatic equation of state with index $\Gamma$.
This combination of parameters results in a disk whose aspect ratio $H/R \simeq 0.06$--0.1
over its entire radial extent when $H$ is defined as $\sqrt{2}c_{s0}/\Omega$, for $c_{s0}$
the isothermal sound speed.
Equivalently, the angle subtended by one scale-height is $\simeq 4^\circ$--$6^\circ$.
The initial magnetic field in the disk is a set of nested poloidal field loops
defined by the vector potential
\begin{equation}
A_{\phi} = \rho - \rho_{C},
\end{equation}  
where $\rho_{C} = 0.1$ and $\vb = \nabla \times \vec{A}$.  The field is scaled so that the
volume-integrated ratio of the gas to magnetic pressure, is initially $25$.

At the beginning of the MHD simulation (which we call \runm), we perturb the pressure
with random fluctuations
whose rms amplitude is $\simeq 1\%$.  From these perturbations, the magneto-rotational
instability grows, and we follow its development {\it without any external torques} for
15 orbits at $r_{\rm M}$.  At this point, the MHD turbulence is fully saturated.  In
addition, the internal stresses due to the anisotropy of the turbulence have led to
significant disk spreading.  Roughly a third of the initial disk mass is lost, mostly
via accretion through the inner boundary of the simulation (at $r = 4$).  In addition,
dissipation associated with the artificial bulk viscosity necessary to describe shocks
properly has heated the gas so that $H/R \simeq 0.12$--0.2 across most of its extent.

The Lense-Thirring torque is turned on only at this point, when the MHD turbulence has
saturated.   For reasons we explain momentarily, we choose the spin-axis of the central mass
to lie two initial scale-heights ($12^{\circ}$) away from the initial orbital axis.  We set the
magnitude of this torque so that $\omega(r_{\rm M})/\Omega(r_{\rm M}) = 1/15$
for precession frequency $\omega$ and orbital frequency $\Omega$.  In terms of the
exigencies of simulation, this is a very natural choice: $\omega \ll \Omega$
through all the disk except its innermost rings, but $\omega$ is not so small that
to follow a precession period would take a prohibitively large amount of computer
time.  Regrettably, in terms of actual physics this is a very unnatural (and nominally
inconsistent) choice because such a ratio is achieved in real life only when
$r/r_g = \{6.1(a/M)\sin\theta[1 + \sqrt{1 + 0.133/\sin\theta}]\}^{2/3}$, where
$\theta$ is the inclination angle between the orbital angular momentum and the central mass's
spin angular momentum.  Newtonian dynamics are, of course, a very poor approximation
in such a location.  We justify it, however, on the grounds that
what is most important to our effort to elucidate disk response to Lense-Thirring torque
is the ability to explore the consequences of significant precession while assuring that it is still
significantly slower than the orbital frequency.    For similar reasons,
we also ignore the relativistic contribution to the apsidal precession rate, even though
one contribution to it is independent of black hole spin and larger than the Lense-Thirring rate,
while another contribution is comparable to the Lense-Thirring precession frequency. 
There are also two more reasons to ignore the relativistic apsidal precession.   It is not the
only mechanism causing the radial epicyclic frequency to differ from the
orbital frequency; radial pressure gradients also do this, and are likely to be at least
as large, especially at the large distance from the black hole at which the Bardeen-Petterson
transition radius is most likely to fall in real disks.  In addition, because only part of the apsidal
precession is proportional to
black hole spin, our Newtonian approximation makes it impossible to scale the apsidal
precession frequency with the Lense-Thirring (nodal) precession frequency.

We then follow the evolution of the disk for 15 orbits at $r_{\rm M}$, i.e., one
full precession period at $r_{\rm M}$, or $T = 30\pi/\Omega_M$ (throughout
the remainder of this paper, we will describe time in units of a ``fiducial orbit",
the orbital period at $r_M$). 
Over the course of this torqued phase of evolution, $\simeq 10\%$ of the disk mass is
accreted through the inner radial boundary.  In addition, the heating associated with a large
number of weak shocks increases its scale height by $\simeq 10\%$.

The contrasting hydrodynamic simulation (called \runh) begins from an initial condition
whose radial profiles
of midplane density and midplane scale-height match the azimuthally-averaged values in the
MHD simulation immediately before the torque is turned on.  The vertical density structure
is what would be expected in hydrostatic equilibrium:
\begin{equation}
\rho(R,z) = \rho_{0}(R) \exp \left[ \frac{-z^2}{H^2(R)} \right],
\end{equation}
where $R$ is cylindrical radius and $z$ is vertical distance away from the disk plane.
The local pressure is simply $H^2(R)\Omega^2(R)\rho/2$.  The velocity field is chosen so
that it is Keplerian in the disk midplane, but the disk rotates on cylinders.  Although
the disk so defined is in vertical equilibrium, there are unbalanced radial pressure
gradients, but they are relatively small.

For the hydrodynamic simulation, the torque begins immediately.  Like the MHD simulation,
it is run for 15 orbits at $r_{\rm M}$.

Simulating a warped accretion disk using a grid-based method presents certain challenges.
In a warped disk it is guaranteed that at least some orbital velocities are oblique to the
grid coordinates, and this obliquity of the strongly supersonic flow must cause at least some 
numerical dissipation (see \cite{SKH13} for numerical experiments quantifying
these effects).  We have made
several choices designed to minimize this dissipation.  Following the results of
the numerical experiments in \cite{SKH13}, we adopt a spherical grid.  We also align
the initial orbital plane of \runm\ with the equatorial plane of the coordinates in order
to minimize numerical dissipation during the 15 orbits in which the MHD turbulence
grows.  This is also why we chose a relatively small inclination between the spin-axis and
the initial orbital axis, and, as we are about to discuss, used as fine a resolution as possible.

The spatial domain for both simulations was
$(r,\theta,\phi) \in [4,28] \times \pi [0.2,0.8] \times 2\pi [0,1]$.  Recent work on
convergence in MHD disk simulations \citep{HGK11,Sorathia12} has shown that at least
32 ZPH (Zones Per vertical scale Height) are required to approach convergence in flat
disks.  It is possible that the additional complexity of external torques and disk
warping raise that standard, but no systematic studies yet exist to determine whether
they do and to what degree.  On the other hand, a sufficiently long simulation even
with this minimal resolution requires a large amount of computer time.   We have therefore
chosen to just meet that standard (and our MHD simulation still consumed
$1.3 \times 10^6$~processor-hours).    Our spherical grid had
$(288,384,1024)$ cells in the radial, polar, and azimuthal directions respectively.  In order
to maximize our effective resolution in the regions we most care about, we space the cells
logarithmically in the radial direction (i.e.,
$\Delta r/r$ is constant), uniformly in the azimuthal direction, and employ a polynomial
spacing in the polar dimension (Eqn.~6 of \citet{NKH10}, with $\xi = 0.65$ and $n = 13$).
This sort of polynomial spacing focuses cells near the equatorial plane, giving a resolution
of more than $32$ ZPH within $\pm \indeg{20}$ of the midplane.  At the pressure maximum of
the torus, $r_M = 10$, the cell dimensions in the radial and polar directions are approximately
equal, while the azimuthal cell dimensions are about a factor of 2 larger.  We present
detailed resolution quality data for this simulation in the Appendix. 

The requirements for hydrodynamic resolution are not nearly so demanding because a purely
hydrodynamic simulation is not turbulent.  Although the flow is complicated, it is laminar,
and has rather little structure on small spatial scales.  Consequently, \runh\ is well-resolved
when run on a very similar grid to \runm, but one whose cell dimensions are exactly twice
as large in each dimension.

Boundary conditions for the two simulations are also identical.  For hydrodynamic quantities,
we use zero-gradient extrapolations and enforce an outwardly-directed velocity in the ghost zones.
For the magnetic field, we set the transverse field components in the ghost zones to zero and
require the normal component to satisfy the divergence-free constraint.

\section{Results}
\label{sec:results}

At a qualitative level, the MHD and purely hydrodynamic simulations appear to
resemble one another strongly.  As we will emphasize throughout the remainder
of this paper, the dominant mechanisms underlying the Bardeen-Petterson
effect are {\it hydrodynamic}, not magneto-hydrodynamic.  That being said,
MHD does create an important difference between them whose consequences
have numerous implications: the MHD system is turbulent, whereas the HD
system is laminar.

\subsection{Precession}
\label{sec:precession}

\begin{figure}
\begin{center}
\includegraphics[width=0.6\textwidth,angle=90]{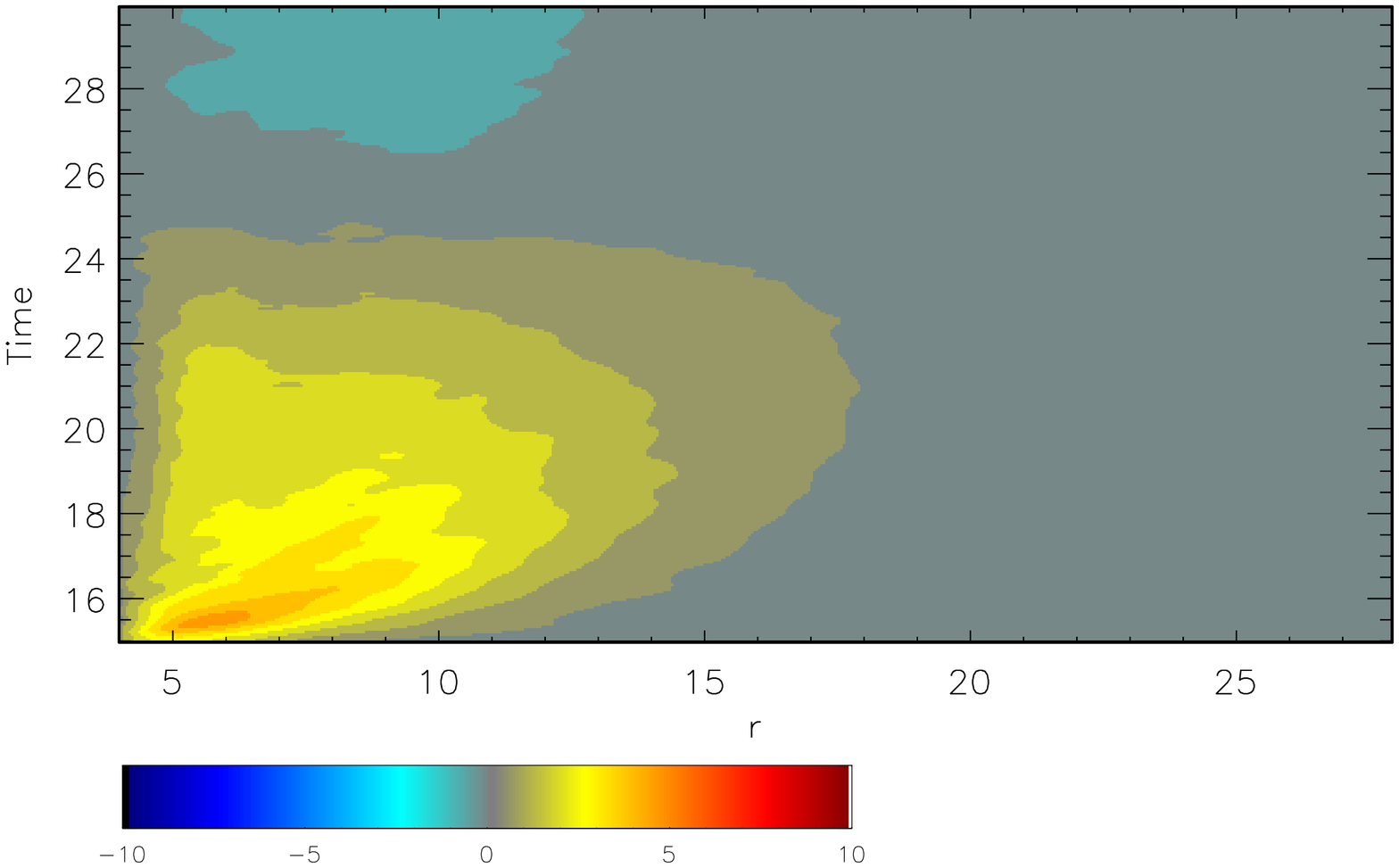} \\
\includegraphics[width=0.6\textwidth,angle=90]{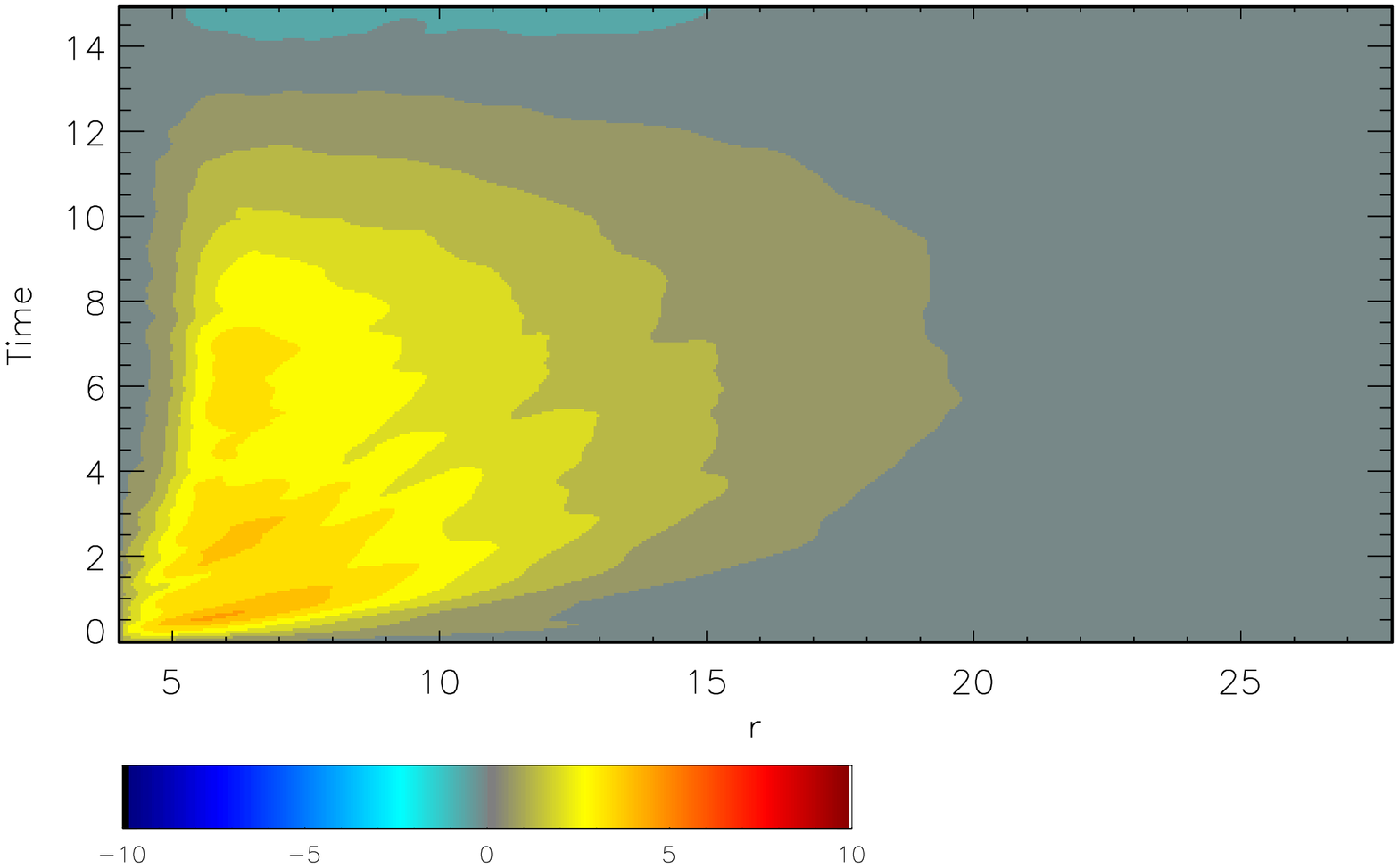} \\
\caption{Color contours (see color bar) of $G_x$ as a function of radius and time.
Upper panel is \runm; lower panel is \runh .}
\label{fig:Gx}
\end{center}
\end{figure}

\begin{figure}
\begin{center}
\includegraphics[width=0.6\textwidth,angle=90]{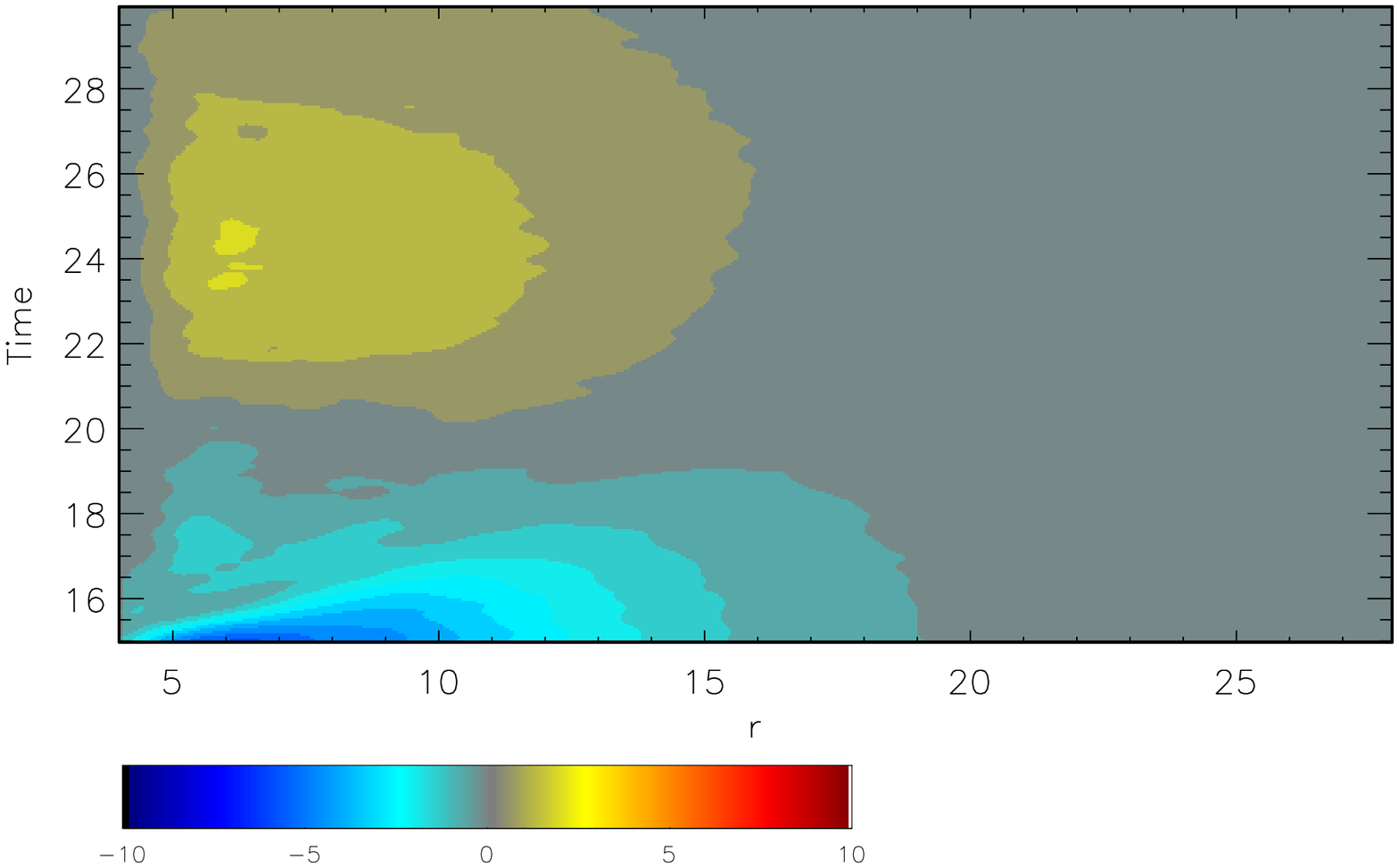} \\
\includegraphics[width=0.6\textwidth,angle=90]{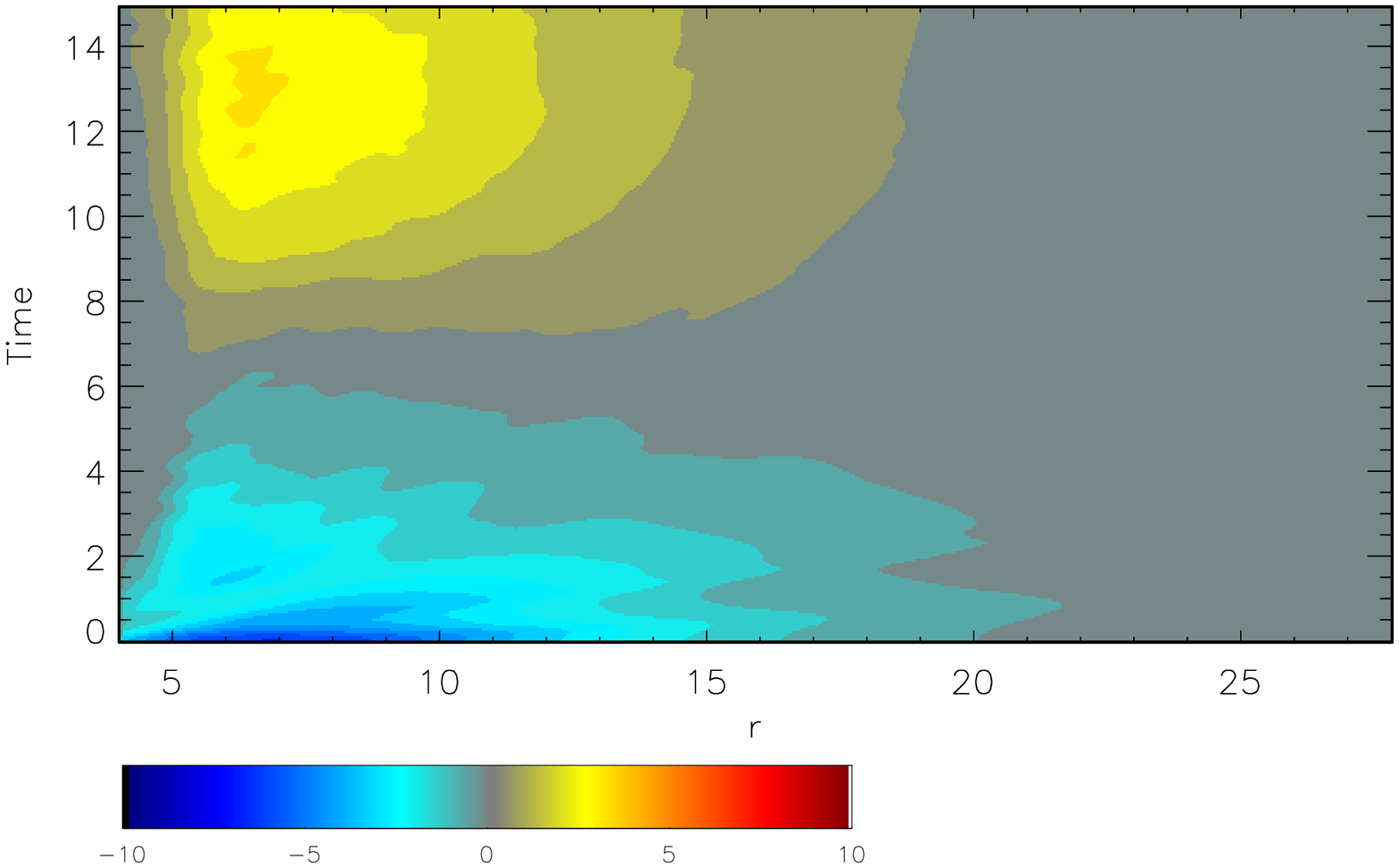} \\
\caption{Color contours (see color bar) of $G_y$ as a function of radius and time.
Upper panel is \runm; lower panel is \runh .}
\label{fig:Gy}
\end{center}
\end{figure}

The ultimate driver of the entire process is Lense-Thirring torque.  To describe it,
as well as the rest of the angular momentum flow, we use a Cartesian coordinate system
$(x,y,z)$ oriented to the central mass's spin axis.  That is, the $z$-direction is defined by the spin
axis.  The $x$ direction in this system is defined so that the
initial disk angular momentum is in the $x$-$z$ plane with $L_x < 0$ and $L_z > 0$.
In terms of these coordinates, the torque ${\bf G}$ has only two non-zero components, $G_x$
and $G_y$.  Their dependence on radius and time is shown in Figures~\ref{fig:Gx}
(the $x$ component) and \ref{fig:Gy} (the $y$ component).

Both runs begin with $G_y < 0$ and relatively strong within $r \simeq 10$.  Likewise
in both cases
$G_y$ passes through zero and changes sign 5--6 orbits after the torque begins,
becoming positive at later times.  The only contrast in this regard is that $G_y$
at late times is rather smaller in the MHD case.   Similarly, a short time
after the torque begins, $G_x$ becomes positive in both cases, particularly for
$r \lesssim 10$, but diminishes in magnitude with increasing time.  A few orbits
before the end of both simulations, $G_x$ also changes sign, a couple of orbits
earlier in \runm\ than \runh.  In both cases, too,
the time-integral of $G_x$ is dominated by $r \lesssim 10$.

\begin{figure}
\begin{center}
\includegraphics[width=0.6\textwidth,angle=90]{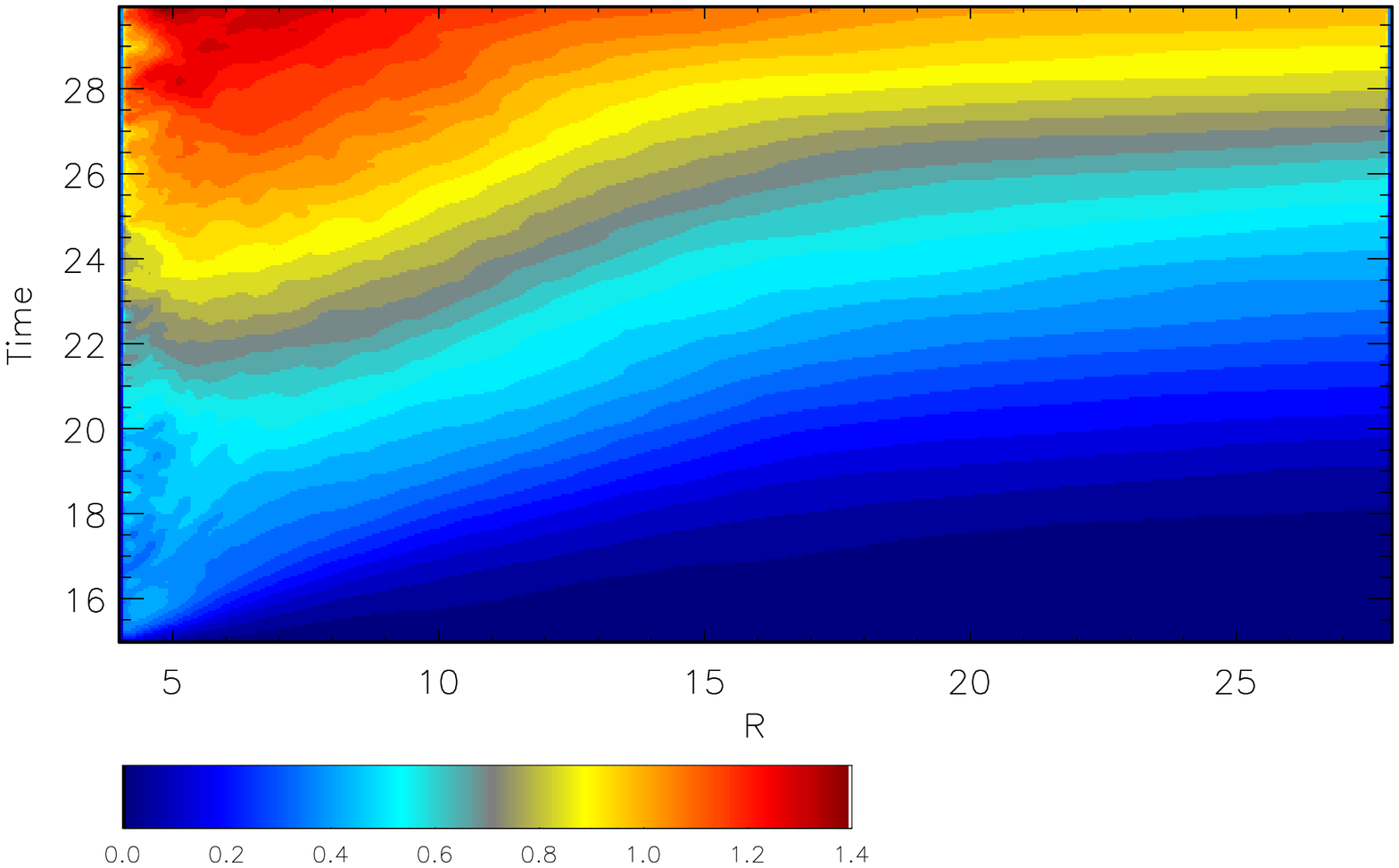} \\
\includegraphics[width=0.6\textwidth,angle=90]{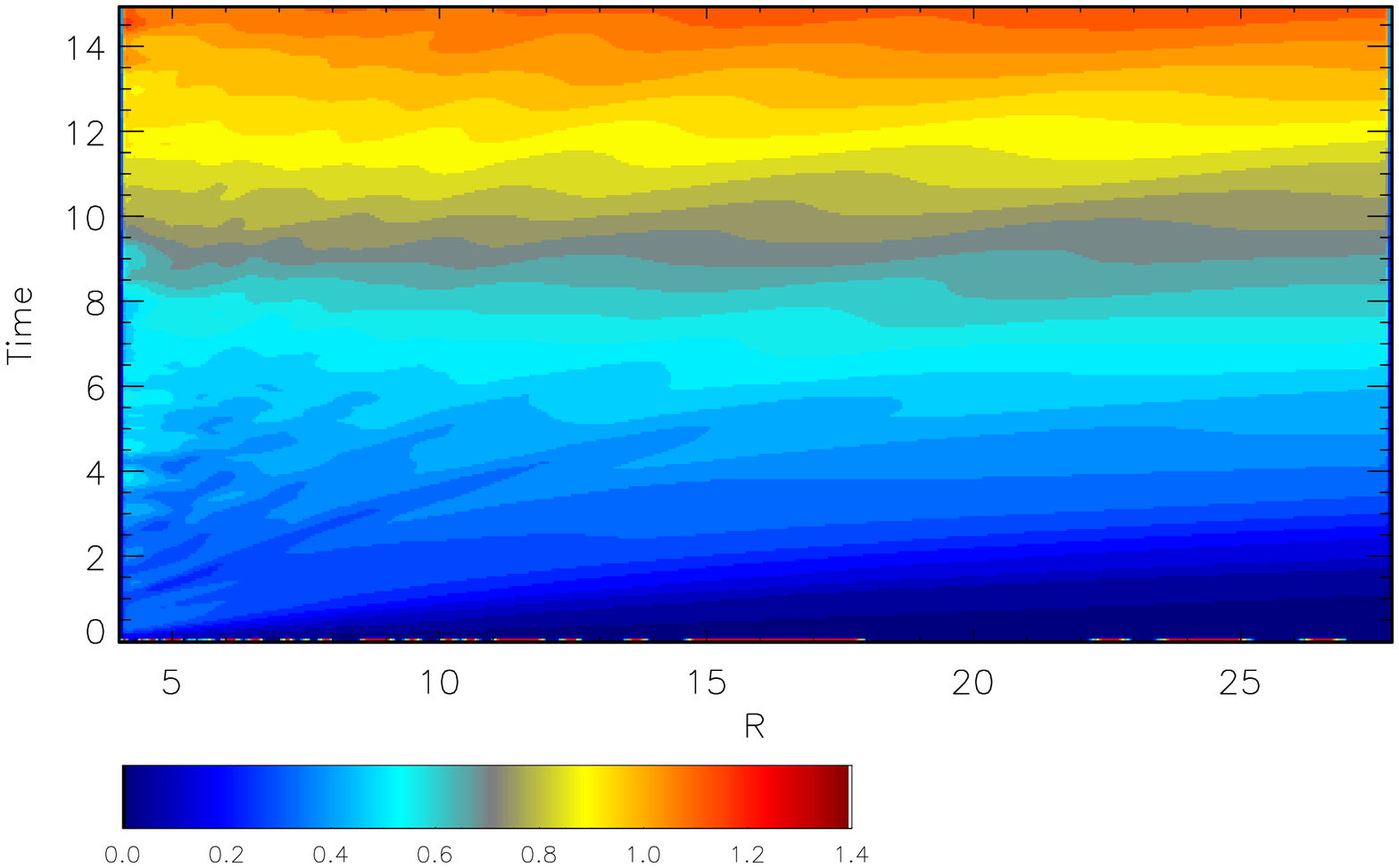} \\
\caption{Color contours (see color bar) of precession angle in units of $\pi$ as a function
of radius and time.  Upper panel is \runm; lower panel is \runh .}
\label{fig:phi_prec}
\end{center}
\end{figure}

These trends reflect the progress of disk precession and alignment.  We measure
the precession angle at radius $r$ by
$\arctan{[(\partial L_y/\partial r)/(\partial L_x/\partial r)]}$,
where the angular momentum partial derivatives are the angular momentum integrated on
spherical shells at radius $r$.  We measure the (mis)alignment angle $\beta$ by
$\arctan{[(\partial L_\perp/\partial r)/(\partial L_z/\partial r)]}$, where $L_\perp^2 =
L_x^2 + L_y^2$.  The sign change in $G_y$
is associated with precession through an angle of $\pi/2$, while the decrease in
magnitude in the torque is a signature of disk alignment.  Figure~\ref{fig:phi_prec} shows the
precession in greater detail.  Like the torque, of course, the precession rate in
the two simulations is overall similar.  However, they are by no means identical.
When MHD effects operate, the mean disk precession is slightly faster than when they are absent.
The largest precession angle at any radius found in \runm\ after 15 orbits is $\simeq 1.4\pi$,
but only $\simeq 1.1\pi$ in \runh.
Especially in the HD case, the precession is not far from solid-body, a result previously
seen in other simulations  \citep{NP00,Fragile05,Fragile08}.
After 2--3 orbits of torque, the color contours for \runh\ run almost flat across the radius--time plane.
Differential precession is weak in an absolute sense in \runm, but nonetheless noticeably stronger
than in \runh.  For the first $\simeq 5$ orbits, compared to \runh\ it precesses more rapidly at small
radii, but more slowly at large.  These contrasts diminish over time.  At the
end of the simulation, the contrast in precession angle across the entire radial span even in
\runm\ is only $\simeq 0.4\pi$,
even though
the precession phase difference between test-particles
at $r=10$ and $r=20$ would have been $15\pi/8$, and between $r=5$ and $r=10$, $15\pi$!
The rate of this approximate solid-body precession corresponds to the test-particle
precession frequency at $r \simeq 11.5$, slightly outside the pressure maximum, and rather
close to the radius corresponding to the mean specific angular momentum of the disk.

On the other hand, although the end-result is nearly solid-body precession, there are
noticeable departures from rigid precession, particularly in \runm
(a possible explanation for why the MHD case is farther from solid-body precession than the
HD case will be presented in \S~\ref{sec:warp}).
As expected, the sense of the contrast is almost monotonic---outer rings precess more
slowly than inner rings.  This sense is not without exception, however---in the inner disk there
can be small departures from monotonicity at the $\simeq 0.1\pi$ level.

Over much of the region where the disk departs from solid-body precession, the slope of the
contours of fixed precession angle is very nearly constant at $\simeq 1.5$ length units
per fiducial orbital period.    Because the disk is close to precessing at a single rate, this
near-constant slope translates to a near-constant twist rate:
$\partial \phi/\partial r \simeq 0.14$~radians per radial length unit.

\subsection{Local warping}
\label{sec:warp}

The degree of local warp can be quantified in terms of
\begin{equation}
\hat\psi \equiv \frac{|d{\bf l}/d\ln r|}{H/r},
\end{equation}
where ${\bf l}(r) \equiv {\bf L}(r)/|{\bf L}(r)|$ is the direction of the angular momentum
${\bf L}$ at radius $r$.
When we compute $\hat\psi$, we use the actual value of $H$ at that location and time.
As shown initially in \cite{NP00} and discussed at greater length
in \cite{SKH13}, the magnitude of this quantity relative to unity gives a good indication of
the degree of nonlinearity of the warp.  That is, the radial contrast in pressure across
a distance $\sim r$ becomes order unity when $\hat\psi \sim 1$ so that the speed of the
corresponding radial flow becomes transonic.

\cite{SKH13} found a further significance for $\hat\psi \gtrsim 1$: the
rate at which warps decay as a result of the angular momentum mixing associated with these
radial flows increases sharply when $\hat\psi$ becomes greater than unity.  That finding
also applies to these simulations.  Despite the strong radial-dependence of the precession
frequency, $d{\bf l}/d\ln r$ never exceeds $\simeq 0.6$ in either simulation; test-particle
precession would have made this figure $\simeq 50$ between $r=5$ and $r=10$ by the end
of the simulation.

\begin{figure}
\begin{center}
\includegraphics[width=0.6\textwidth,angle=90]{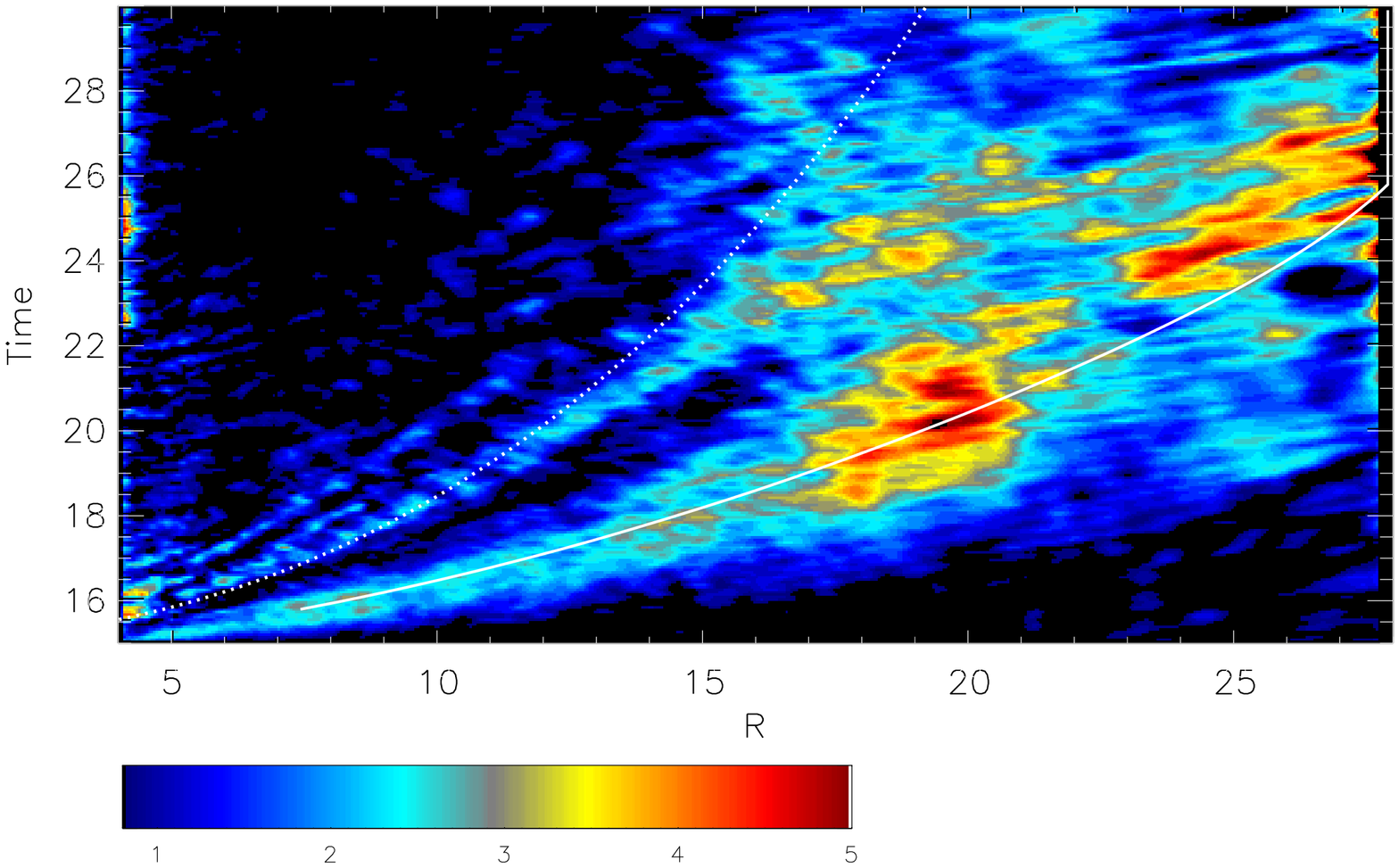} \\
\includegraphics[width=0.6\textwidth,angle=90]{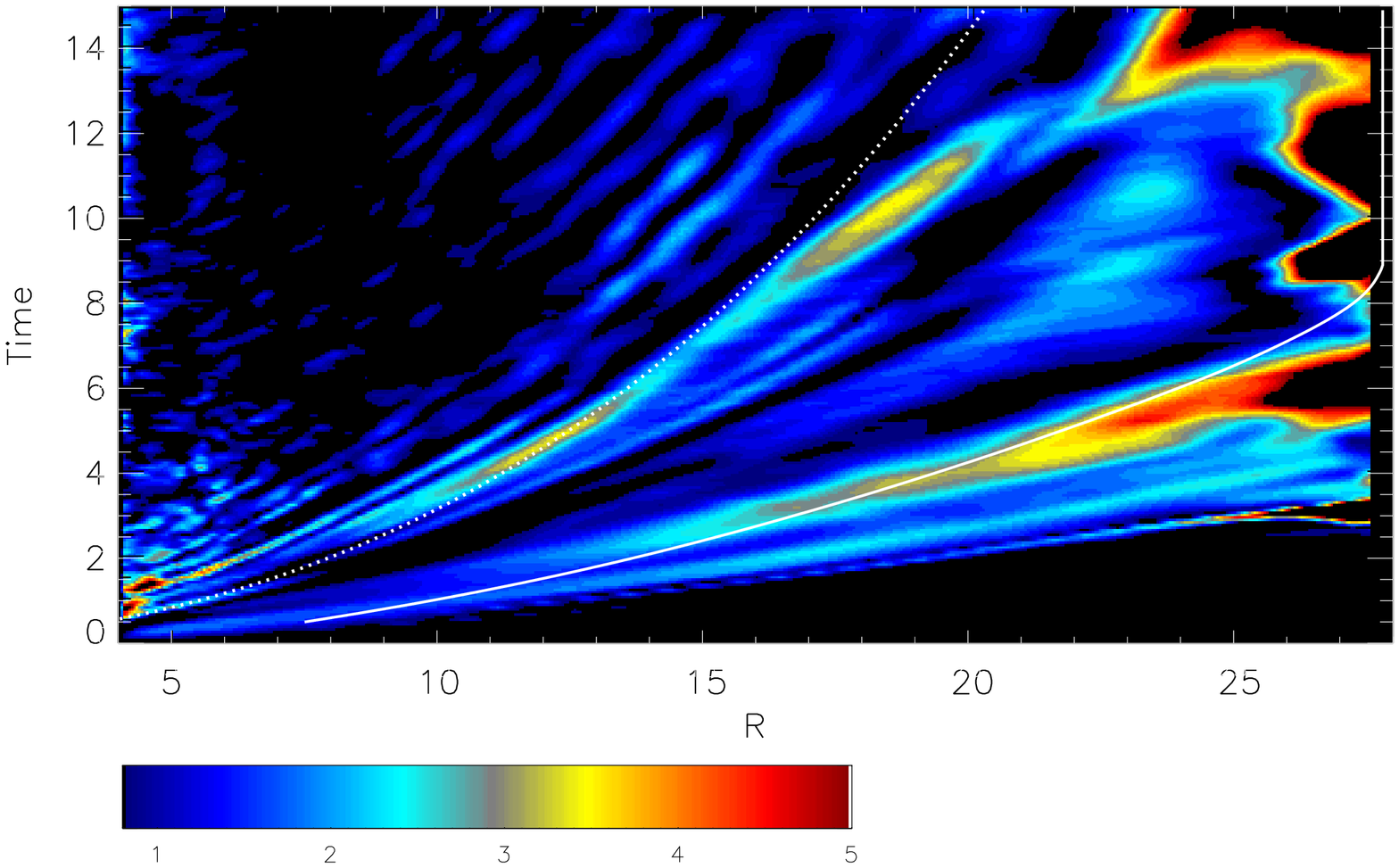} \\
\caption{Color contours (see color bar) of $\hat\psi$.  Upper panel is \runm; lower panel is
\runh .  The color scale goes to black for $\hat\psi \leq 0.8$ in order to emphasize the 
boundary between linear and nonlinear warps.  In each panel, there are two superposed
white curves.  The solid one represents the trajectory of a bending wave (with speed
one half the mass-weighted isothermal sound speed: \cite{Lubow02}), the dotted one shows
the trajectory implied by eqn.~\ref{eqn:precessray} with $t_0(r)$ as described in the text. }
\label{fig:psihat}
\end{center}
\end{figure}

Figure~\ref{fig:psihat} shows how the warp parameter varies as a function of radius
and time in both simulations.  One way to view this pair of figures is to focus on
behavior as a function
of time at a fixed radius.  From this perspective, we see that $\hat\psi$ oscillates
between quite small and $\simeq 2.5$ on timescales of order a fiducial orbit.
In other words, the warp induced by the radially-varying external torque appears
to exhibit a ``stick-slip" behavior: the differential torquing builds the warp until
$\hat\psi$ reaches this maximum value, and then the strong radial flows associated with
such a nonlinear warp rapidly erase it.  The warp grows larger at the outside of
the disk, where the surface density is small.  It is also noteworthy that the ``stick-slip"
oscillation is considerably more sharply defined in the HD case than in the MHD case;
this is not surprising in view of the turbulence that is the hallmark of the latter,
but nonexistent in the former.

The locations of strong warping propagate coherently outward over time.  Although the
correspondence is not perfect, the trajectory of the first and strongest pulse in both
simulations is close to what would be expected for a bending wave.   The time-width
of this pulse defines a characteristic frequency, $\omega_* \sim 2$ radians per fiducial
period.  Bending waves more than a few times greater than the local precession
frequency travel at half the local isothermal sound speed; bending waves
with lower frequencies travel more slowly, becoming  non-propagating when their
frequency drops below the local precession frequency \citep{Lubow02}.
Because the precession frequency reaches $\omega_*$ only for $r \lesssim 7$
and decreases rapidly outward ($\propto r^{-3}$), the asymptotic wave speed applies
to most of the mode content for these pulses for all radii $\gtrsim 7$.   We plot tracks
defined by this speed in Figure~\ref{fig:psihat},
where it appears to provide a reasonable approximation to the propagation of the
first pulse in both \runm\ and \runh\ .    That the first pulse in \runh\ spreads slowly
with radius indicates a spread in propagation speeds, suggesting
the presence of wave components with frequencies ranging down from $\omega_*$
to $\gtrsim \omega(r)$.

In \runh, but not in \runm, there is also a rough correlation between the trajectory of that
first pulse and the establishment of approximately solid-body precession.   A clue to the
origin of this contrasting behavior may be found in the much more irregular variation of warp
magnitude along the trajectory of the pulse in \runm\ than in \runh.   We suggest that
the turbulence permeating the disk in \runm\ disrupts smooth propagation of such a wave: the ratio
$\left(\langle v_{Ar}v_{A\phi}\rangle/c_s^2\right)^{1/2} \simeq 0.2$--0.3 (the
average is over spherical shells) immediately before
the torques begin (here $v_{A(r,\phi)} \equiv B_{(r,\phi)}/\sqrt{4\pi\rho}$).
Laminar magnetic field would probably be less effective
in interfering with a bending wave because the magnetic tension on such long length
scales (vertical wavelength $\simeq 2H$) is relatively weak; for example, the growth
rate of an MRI mode with the vertical wavelength of the bending wave is only $\sim \Omega/7$.
That solid-body precession is never quite achieved in \runm\ will prove crucial in our
analysis of the contrasting alignment behavior shown by these two simulations.

Later $\phat$ pulses, however, propagate substantially more slowly and decelerate
outward.   In fact, their speeds decrease steadily from each late pulse to the next.
These observations suggest that these pulses are not driven by bending wave
dynamics.  As we have just seen, the speeds of bending waves are controlled
by the isothermal sound speed and the relationship between their frequency and
the local precession frequency.    Neither the sound speed nor the local precession
frequency is a function of time, while the time-dependence of the pulse widths
suggests that $\omega_*$ increases slowly.    Thus, the speeds of bending
waves should vary little with time or possibly become even closer to the asymptotic
value of half the isothermal sound speed, whereas the speeds of these pulses
become progressively slower and slower at later times. 
Instead, the later pulses appear to be better
described by differential precession twisting the disk from flat to a critical value of
$\hat\psi$.  In this essentially kinematic picture, the disk begins in a state in which
it is locally flat (more precisely, it is flat near radius $r$ at time $t_0(r)$).
The differential torques then build a warp without (at first) any coupling between
adjacent rings of gas.  Once the warp grows to the point at which $\hat\psi = \hat\psi_{\rm crit}$,
neighboring rings couple through radial flows and, after about one local orbit, that
region of the disk is once again flat.  In this picture, the radius $r_{\rm crit}$ at
which $\hat\psi = \hat\psi_{\rm crit}$ moves outward in time according to
\begin{equation}\label{eqn:precessray}
r_{\rm crit} = r_M\left\{\frac{6\pi}{H/r}\frac{\omega(r_M)}{\Omega(r_M)} 
       \frac{\sin\beta}{\hat\psi_{\rm crit}} \, \left[t - t_0(r)\right\} \right]^{1/3},
\end{equation}
where $\beta$ is the angle between the orbital axis and the central mass's spin-axis, and
time is measured in fiducial periods (orbital periods at $r_M$).  The dotted curve in
Figure~\ref{fig:psihat} assumes $\hat\psi_{\rm crit} \simeq 2.5$, consistent with the most
common value of $\hat\psi$ along these ridgelines and sets $t_0(r)$ to one local orbit
after the peak warp induced by the bending wave passes that radius.  The delay of
one orbit is a rough approximation to the time required for warp relaxation from
a $\phat$ of that magnitude \citep{SKH13}.  In both cases, but especially for \runh,
this model does a fairly good job of reproducing the track followed by the second
major pulse in $\phat$.  It appears, therefore, that although the twist induced
by the differential torques is initially propagated outward by a bending wave,
subsequent twists---which also have considerably smaller amplitude---propagate
purely kinematically.

\subsection{Alignment}

\begin{figure}
\begin{center}
\includegraphics[width=0.6\textwidth,angle=90]{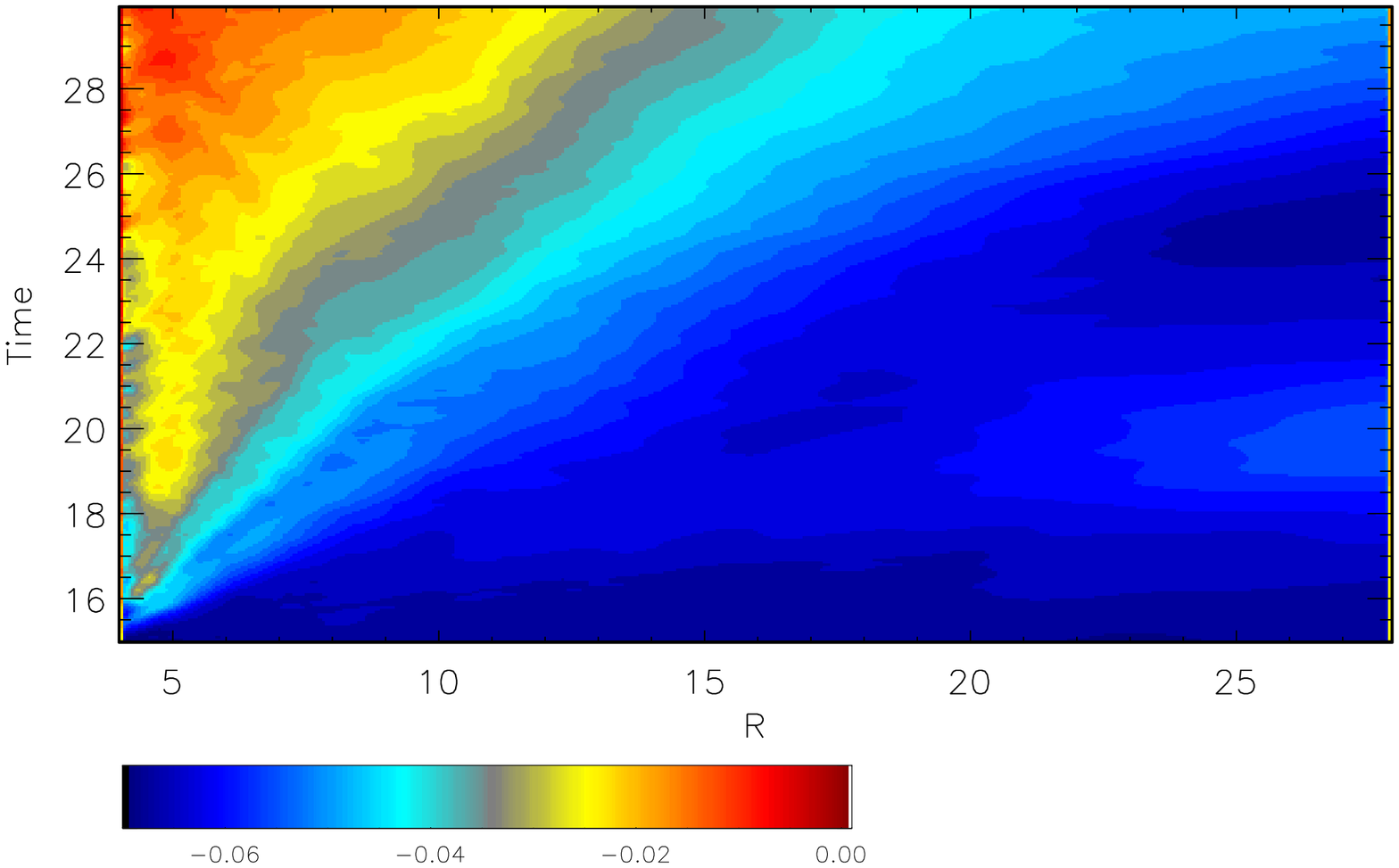} \\
\includegraphics[width=0.6\textwidth,angle=90]{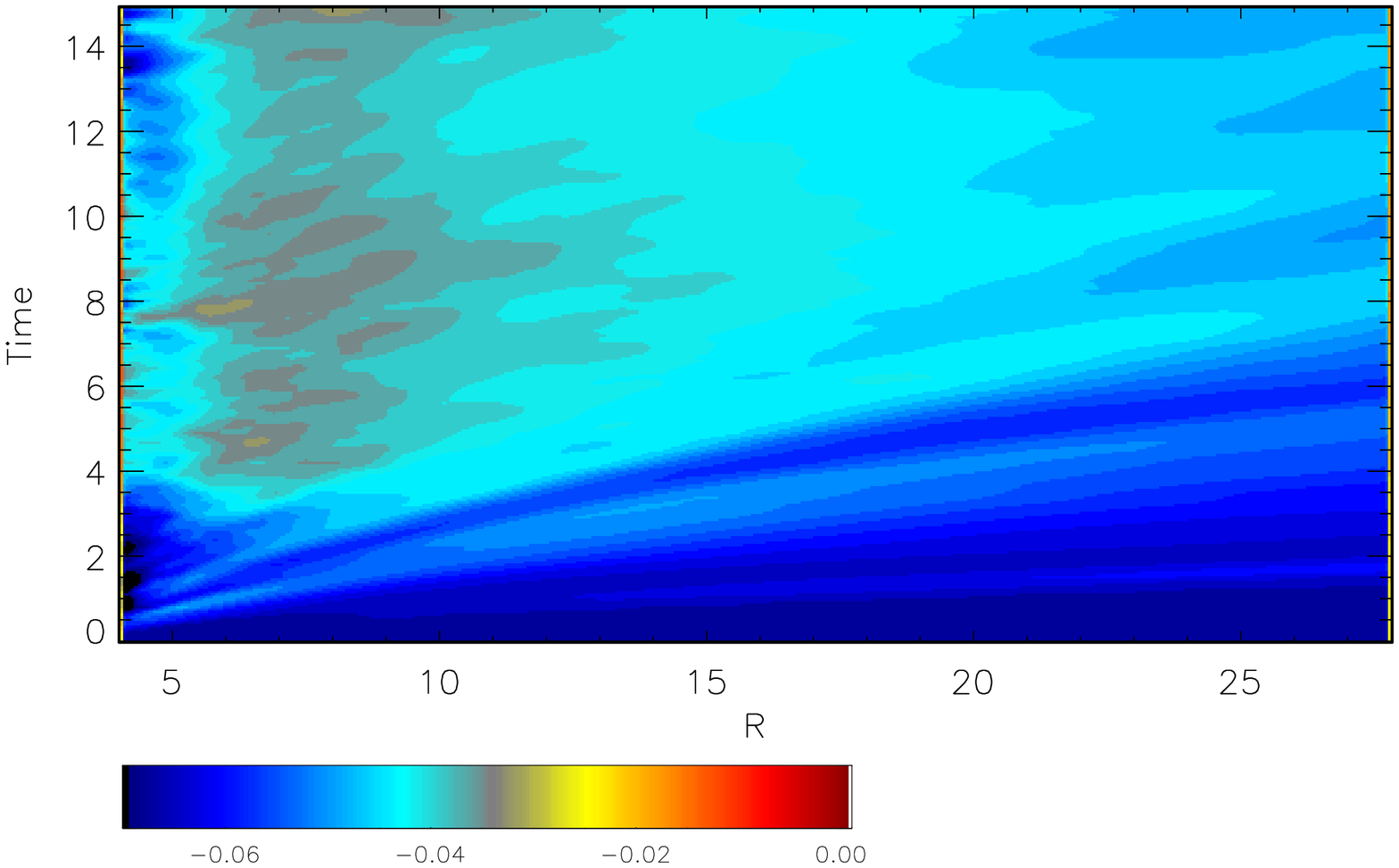} \\
\caption{Color contours (see color bar) of the inclination angle (in units of $\pi$)
as a function of radius and time.
Upper panel is \runm; lower panel is \runh .}
\label{fig:align}
\end{center}
\end{figure}

Figure~\ref{fig:align} shows the alignment angle $\beta$ (in units of $\pi$)
in the two simulations as a function of
radius and time.  Alignment is the respect in which MHD makes the greatest difference.
For the first several orbits, \runh\ aligns significantly more quickly
than \runm\ at large radii, while at small radii the opposite is true.  At later times, however,
alignment virtually stops in \runh\ , while continuing steadily in \runm.  As a
result, whereas half the initial alignment at $r=7$ has been eliminated
after $\simeq 5$~orbits in \runm, that degree of alignment is not achieved even by
the end of 15~orbits in \runh; \runh\ diminishes its misalignment by $\simeq 40\%$
at $6 < r < 9$ by $\simeq 4$~orbits, and then ceases to change alignment thereafter.
By contrast, the entire range of radii interior to $r \simeq 15$ in \runm\ has diminished
its misalignment by at least half by the end of its 15~orbits, and the misalignment has
been sanded down to $< 0.02\pi$ for all $r < 11$.

Comparison of Figures~\ref{fig:align} and \ref{fig:psihat} also shows a close
correspondence between the regions where the inclination angle has been reduced
to less than $\simeq H/r$ and regions where the warp is always in the linear
regime.  This is, of course, a natural consequence of the fact that when the
inclination is $< H/r$, there cannot be radial contrasts in inclination or
orientation any larger than that.   What is more noteworthy about this region
of permanently linear warp is that it is also the region where the inclination
in \runm\ continues to decline, whereas no such improvement in alignment
occurs in \runh.  We will return to this point later.

\subsection{MHD vs. HD}\label{sec:mhdvshd}

\begin{figure}
\begin{center}
\includegraphics[width=0.35\textwidth,angle=90]{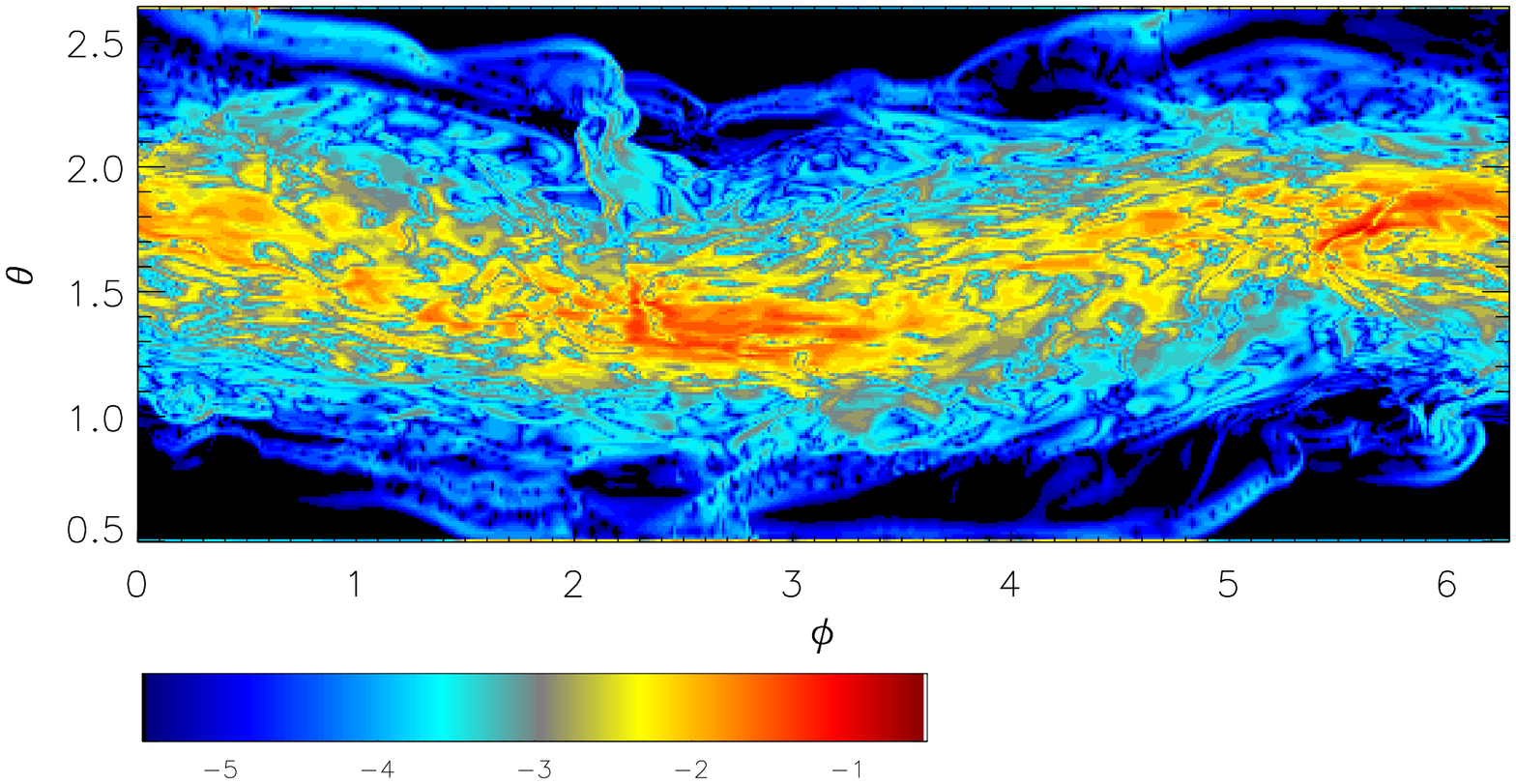} \\
\includegraphics[width=0.35\textwidth,angle=90]{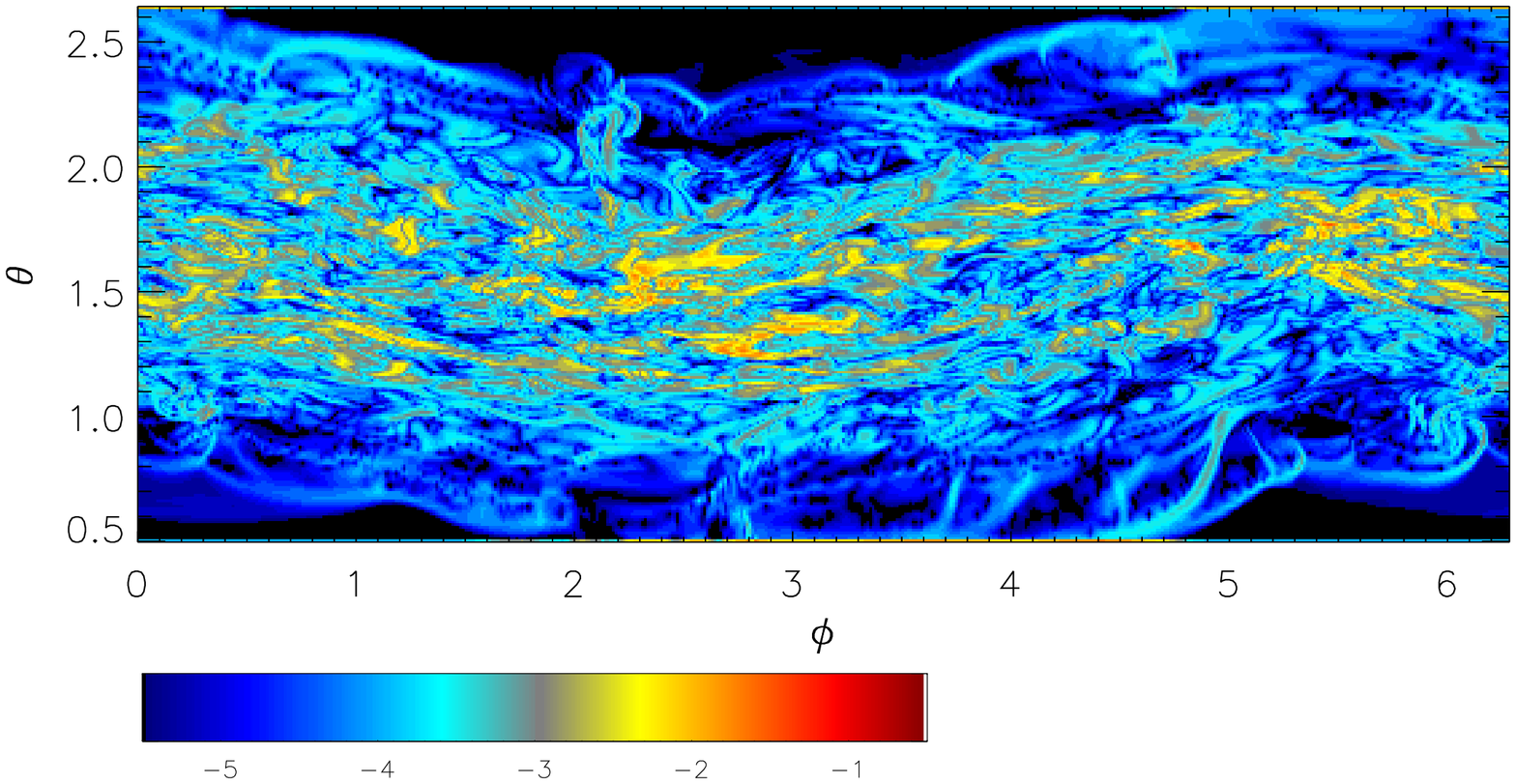} \\
\includegraphics[width=0.35\textwidth,angle=90]{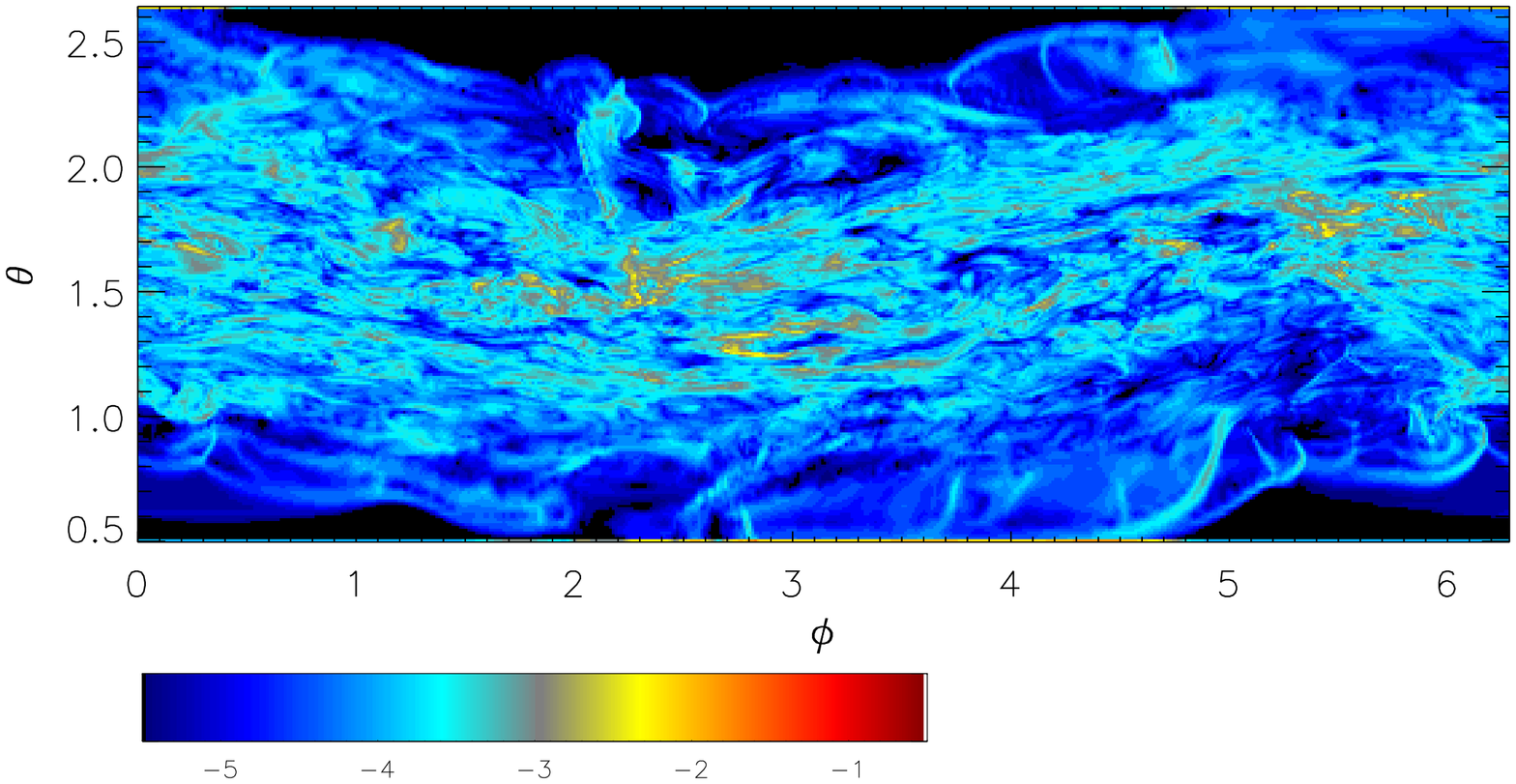} \\
\caption{Color contours on a logarithmic scale of the magnitudes of three force densities:
the radial component of the gas pressure gradient (top panel), the radial component of
the magnetic pressure gradient (middle panel), and the total
magnetic tension (bottom panel).  All are measured as a function of $\phi$ and $\theta$
on the $r=10$ shell at 4 orbits after the torque was turned on in \runm .}
\label{fig:forces}
\end{center}
\end{figure}

As we have already pointed out, at least through the initial stages of alignment,
MHD effects appear to be secondary to hydrodynamic effects, although the sense of
that secondary contribution is to promote alignment.  The data shown in
Figure~\ref{fig:forces} illustrate explicitly the relative importance of
magnetic and pressure forces.  The disk curves up and down in these coordinates
because at this radius and time it has already moved out of the equatorial plane
of the grid.  Compared at the same location during the time when the disk is
aligning most rapidly, the radial gas pressure gradient is generally
$\sim 10$--100 times larger than the radial magnetic pressure gradient, while the
radial magnetic pressure gradient is $\sim 3$ -- 10 times larger than the total magnetic
tension force.  At later times, the magnetic forces rise relative to the fluid
forces, but only by a factor of 2--3.  Thus, in terms of instantaneous forces, magnetic
effects are always considerably weaker than hydrodynamic forces.

The relative weakness of magnetic forces is enhanced by the fact that
after the torque begins, the total magnetic energy in the disk declines sharply,
falling by about a factor of 2 over the first 5 orbits of torque.  From then
until the end of the simulation, the total magnetic energy varies hardly at all.
It is possible that some of the field loss is a numerical artifact, caused by a
combination of newly-created flows oblique to the coordinate grid and an artificially
large rate of magnetic reconnection as the radial flows driven by the disk warps
push regions of oppositely-directed field toward one another.  However, we believe
that it is not entirely artificial.  We have several reasons to think so.  The
first is that the degree of obliquity is never terribly large: an inclination
of $12^\circ$ is not very great, and the numerical diffusion experiments of
\cite{SKH13} found that even with a grid a factor of 4 coarser than ours, only
$\simeq 0.5\%$ of the angular momentum was lost after 10 orbits of integration
at a slightly greater inclination ($15^\circ$).  The second is that before carrying
out \runm, we ran the same problem on a grid a factor of 2 coarser in each
dimension.  The magnetic energy loss after the initiation of the torque in
that run was larger, but not by much, a factor of 3 decrease rather than
a factor of 2.  The third reason is that the radial flows, which sometimes
lead to shocks, do contain regions where reconnection is driven by the
fluid motions; in these cases, the local rate of reconnection may be resolution-dependent,
but its end-result is not.  Thus, some of the field loss we see is likely
due to lack of resolution, but it probably does not account for the entire effect.
It is also possible that the development of the magneto-rotational instability
is altered in the presence of a warp.

Thus, although our simulation includes a full treatment of MHD turbulence, it turns
out to have a relatively small effect on the magnitude of the radial flows primarily
responsible for transporting misaligned angular momentum through the disk.   The
situation in even a magnetized disk in fact resembles quite closely that explored
in \cite{SKH13}, in which the relaxation of disk warps in a purely hydrodynamic
context was studied.   Just as was found in that paper, we find that when $\hat\psi > 1$,
which is the generic situation when there is any substantial inclination, the magnitudes
of these radial flows are primarily controlled by the fluid dynamics of order unity
pressure contrasts in an orbital setting, i.e., quasi-free expansion limited by orbital
mechanics.

Despite the dominance of hydrodynamic effects over magnetohydrodynamic effects
in most aspects of warped disk evolution, we have also seen that MHD appears both
to accelerate alignment and to continue it longer.  It is noteworthy that
the purely hydrodynamic disk ceases alignment progress when its inclination reaches
$\simeq 6^\circ$, here one scale-height (Fig.~\ref{fig:align}b).  At that point
(see also Fig.~\ref{fig:psihat}b), it becomes almost impossible for any warps
to reach nonlinear amplitude.  Consequently, the Reynolds stress responsible
for radial mixing of unaligned angular momentum drops rapidly because it
is a strongly increasing function of $\hat\psi$ when $\hat\psi \simeq 1$
\citep{SKH13}.  On the other hand, when MHD effects are present, the turbulence
they cause creates much short lengthscale structure in the velocity field,
enhancing the angular momentum mixing rate.  This mixing rate is considerably faster than
the inflow rate associated with Reynolds stresses in a flat disk because the
scale of the gradients is much smaller: $\sim 0.1r$ rather than $\sim r$.

We close this section with an examination of an assumption frequently made in other
studies of warped disks: that the vertical shear of radial motions induces a stress that
can be phenomenologically modeled as an isotropic ``$\alpha$ viscosity" \citep{PP83}.   Such
a viscosity would create a viscous stress proportional to the shear (but, of course, with
opposite sign) whose magnitude is $\sim \alpha p$ when $\partial v_r/\partial z \sim \Omega$. 
Some support was given to this hypothesis by \cite{Tork2000}, who measured the
decay of epicyclic motions in a numerical simulation of a vertically-stratified shearing box
with MHD turbulence, although their conclusions were somewhat clouded by the limitations
of their approximations and by their finding that larger amplitude motions were primarily
damped by a different mechanism, the excitation of inertial waves.

\begin{figure}
\begin{center}
\includegraphics[width=0.6\textwidth,angle=90]{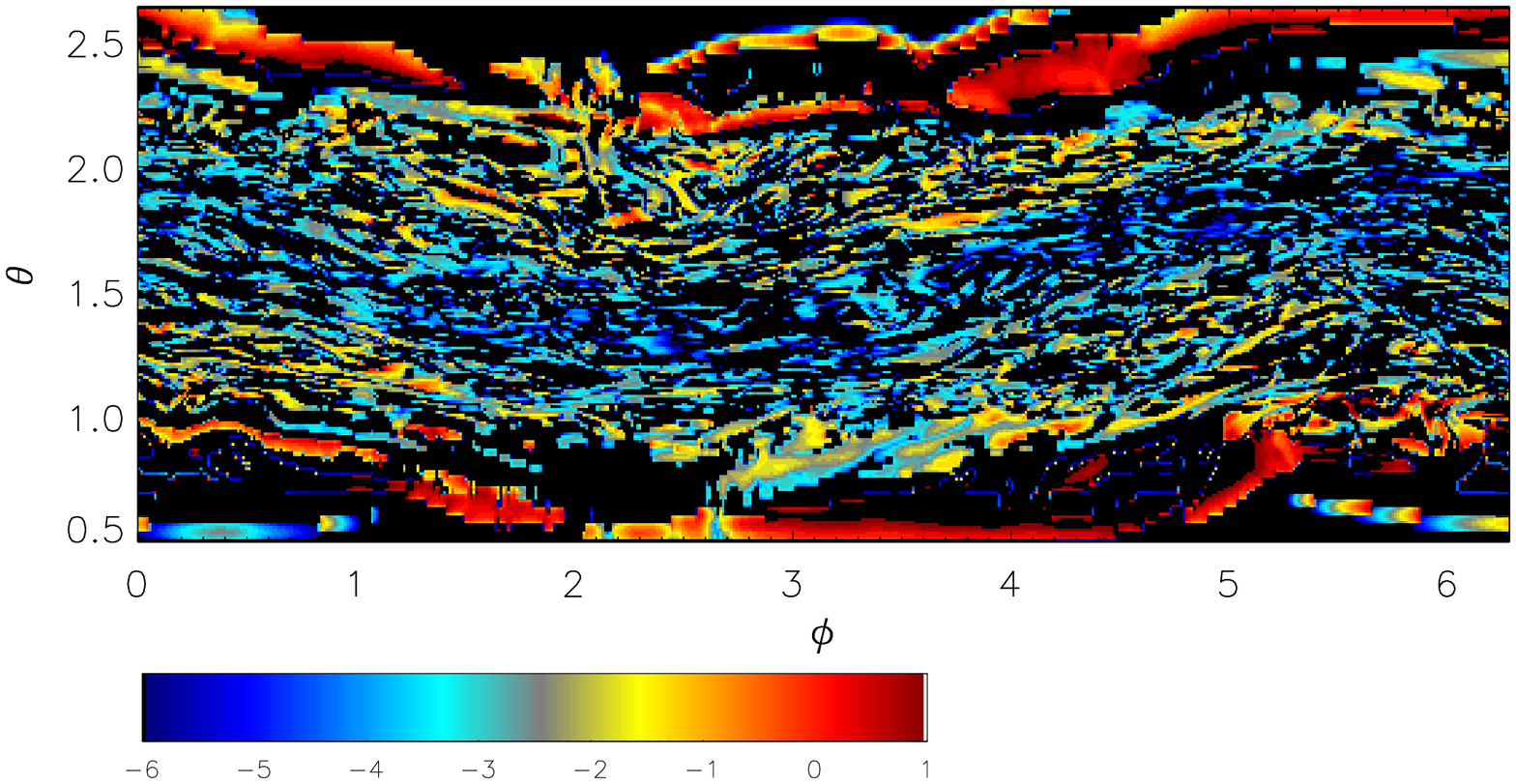} \\
\includegraphics[width=0.6\textwidth,angle=90]{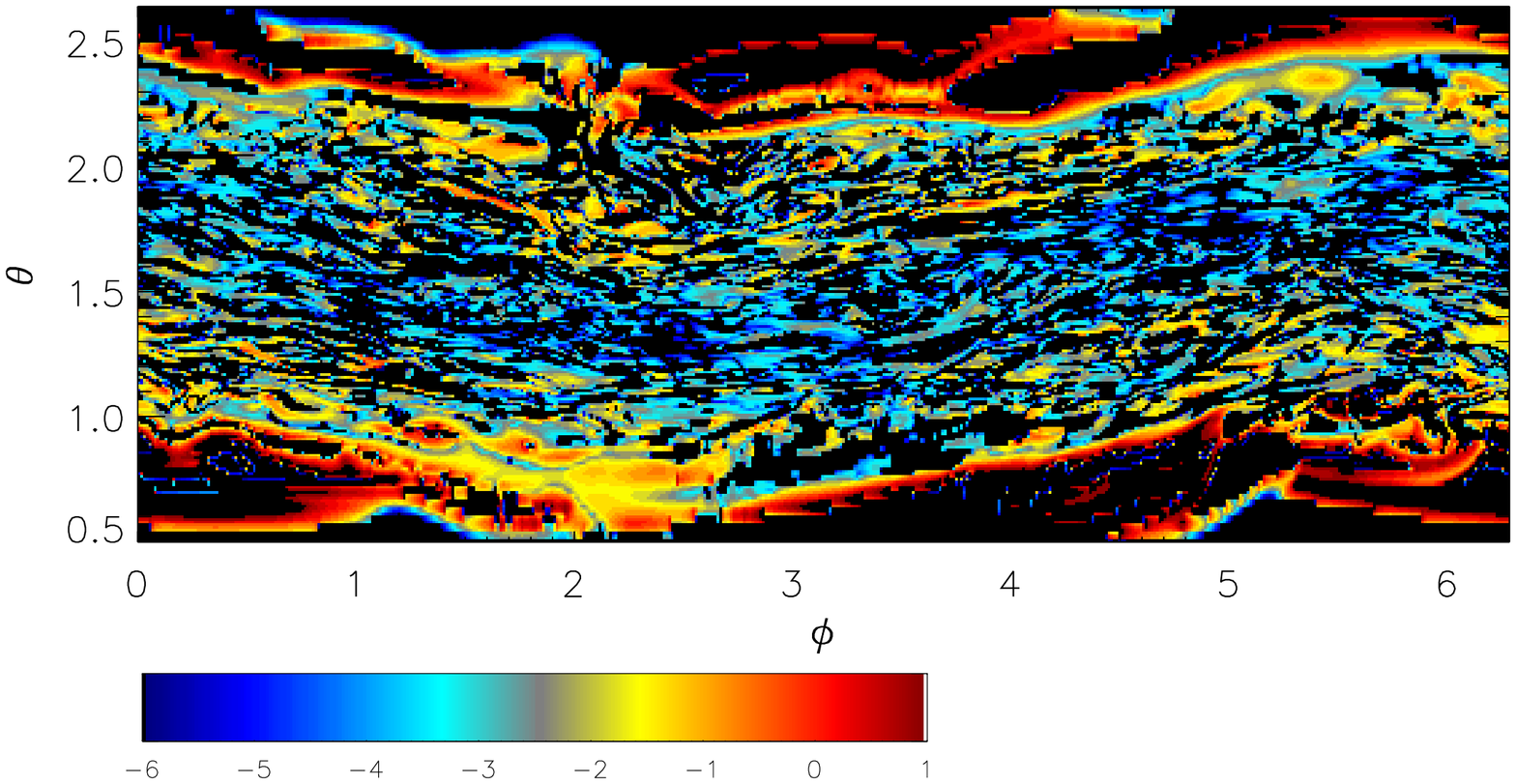} \\
\caption{Color contours (see color bar) of $\log_{10}(|\alpha_*|)$ as defined in text, measured
on the same spherical shell and at the same time in \runm\ as in Fig.~\ref{fig:forces}.
In order to display these quantities on a logarithmic scale,  we separate the data
into two cases: where $\alpha_* > 0$ (upper panel) and where $\alpha_* < 0$ (lower panel).
Black indicates a region where $\alpha_*$ has the opposite sign from the colored data in
that panel.}
\label{fig:alpha}
\end{center}
\end{figure}

Here we test a form of this hypothesis: that the Maxwell stress of MHD turbulence
(whose $r$-$\phi$ component is responsible for accretion) acts in the same manner
independent of the orientation of the shear.  In this context, the relevant component of
the Maxwell stress is $r$-$\theta$.    We therefore compute the ratio
\begin{equation}
\alpha_* = {B_r B_\theta \Omega \over 4\pi p \partial v_r/ \partial z}
\end{equation}
on a sample spherical shell when alignment is beginning at that location.
\footnote{Because of the disk's warp and twist, the $\theta$-direction is exactly
normal to the disk only at large radii where there has been little precession or alignment.
However, the relatively small initial misalignment angle ($12^\circ$) means the
error due to imprecise identification of the disk normal is quite small compared to
the magnitude of the effect we measure.}
If the isotropic $\alpha$ viscosity hypothesis were true, $\alpha_*$ would be consistently
positive and have magnitude $\sim 0.01$--0.1, similar to the ratio of the time-averaged
and vertically-integrated $r$-$\phi$ component of the magnetic stress to the similarly
averaged and integrated pressure.   As Figure~\ref{fig:alpha} demonstrates quite
clearly, neither of these expectations is confirmed.    The quantity $\alpha_*$ is equally
likely to be positive or negative, and its absolute magnitude in the disk
body---including where the shear and Reynolds stress are greatest---is generally $\sim 10^{-5}$--$10^{-4}$.
The mass-weighted mean $\langle \alpha_* \rangle \simeq 3 \times 10^{-5}$.
When averaged over snapshots spanning 0.3--1 fiducial orbit (which is also
the local orbital period for the data shown in Fig.~\ref{fig:alpha}), the magnitude of the shear
decreases, but the radial pressure gradients induced by the warp preserve some overall consistency.
Consequently, the shear diminishes only somewhat when averaged over short time intervals.
On the other hand, time-averaging over even as brief a time as half a fiducial orbit (five
snapshots) reduces the magnitude of the $r$-$z$ magnetic stress by almost an order of magnitude.

To quantitatively calibrate the measured magnitude of $\alpha_*$, we note that the rms magnitude
of the $r$-$z$ shear is
$\simeq 2.6\Omega$ when weighted by mass and $\simeq 6.9\Omega$ when weighted
by volume.   A further sense of scale may be gleaned from the fact that both the $r$-$r$
and the $r$-$z$ components of the Reynolds stress are  frequently $\sim (1$--$10)p$
(and the $r$-$z$ Reynolds stress has no more correlation with the corresponding shear
than the same component of the Maxwell stress).   In other
words, the radial flow speeds are generally transonic, the lengthscale of vertical variation
is several times smaller than the pressure scale height (i.e., turbulence is important),
and the overall dynamics are dominated by pressure gradients and gravity, with only
small contributions from any other sources of stress.

As a final comment, it is worth noting that in several respects this result can be understood
on the basis of qualitative arguments.   First, it is not surprising that the $r$-$z$ component of Maxwell stress
should be much smaller in magnitude than the $r$-$\phi$ component when the magnetic
fields are associated with MRI-driven turbulence.    Simulations of MRI-driven turbulence
in flat disks have consistently found that $|B_\phi| \sim 3|B_r| \sim 10|B_z|$ \citep{hgb,shgb96,HGK11};
even without allowance for the degree of correlation between these components, one
would therefore expect the $r$-$z$ component to be an order of magnitude smaller
than the $r$-$\phi$ component.  Because the consistency of orbital shear imposes a
strong correlation between $B_r$ and $B_\phi$, yet $r$-$z$ shear has no consistent
value, one might also expect that the degree of correlation in $r$-$z$ would be much
weaker than in $r$-$phi$.   The lack of sign correlation can also be understood intuitively.
Magnetic stresses in conducting fluids result from {\it strain} in the fluid, not shear.   If the flow
is oscillatory, strain is $\pi/2$ different in phase from shear, and such a phase offset would entirely
eliminate any sign correlation.  Fluctuations due to turbulence further diminish any direct
tie between magnetic stress and fluid shear.   

\section{Analysis}

\subsection{The alignment rate}\label{sec:align}

To understand these results, it is helpful to frame them in terms of the global angular
momentum budget.
Alignment is often described as due to ``dissipation" associated with angular momentum
diffusion \citep{PP83}.    However, this description is a bit imprecise.   Changing the direction
of an angular momentum first and foremost requires a torque.   The mechanism producing this
torque may or may not be dissipative, and any dissipation involved may or may not be associated with
a process described by a classical diffusion equation.  What is truly essential is that new angular
momentum introduced into the system must be brought to a location where it can cancel the
misaligned angular momentum.
More specifically, there are only three ways the angular momentum of a given disk region
can change: by a divergence of Reynolds stress, a divergence of Maxwell stress,
and an external torque.   However, the Lense-Thirring torque by its very nature
cannot change $|{\mathbf L_\perp}|$ at the location where the torque is exerted
because it is always exactly perpendicular in direction.  It follows that to align a ring
at radius $r$, there must be a way to bring it angular momentum from a
region with a precession phase {\it different} from that of radius $r$, where the
Lense-Thirring torque has a component opposite in direction to ${\mathbf L_\perp}(r)$.
It is possible for diffusive mixing to accomplish this, but it can also be accomplished
by other means, and any mixing process must satisfy certain specific conditions.
The region that is mixed must contain a large enough range of precession phase
that some portion of it has a torque with a direction that can cancel ${\mathbf L_\perp}(r)$,
but not so large that mixing leads to complete cancellation in the net torque.
In addition,  as we have already seen, magnetic forces are
in general quite small compared to hydrodynamic forces, so the Maxwell stress
contributes little to the alignment.   Consequently, alignment must be due to divergences
in the Reynolds stress.  Moreover, because the interesting gradients are all in the
radial direction, it makes sense to think only about radial angular momentum flows.

\begin{figure}
\begin{center}
\includegraphics[width=0.6\textwidth,angle=90]{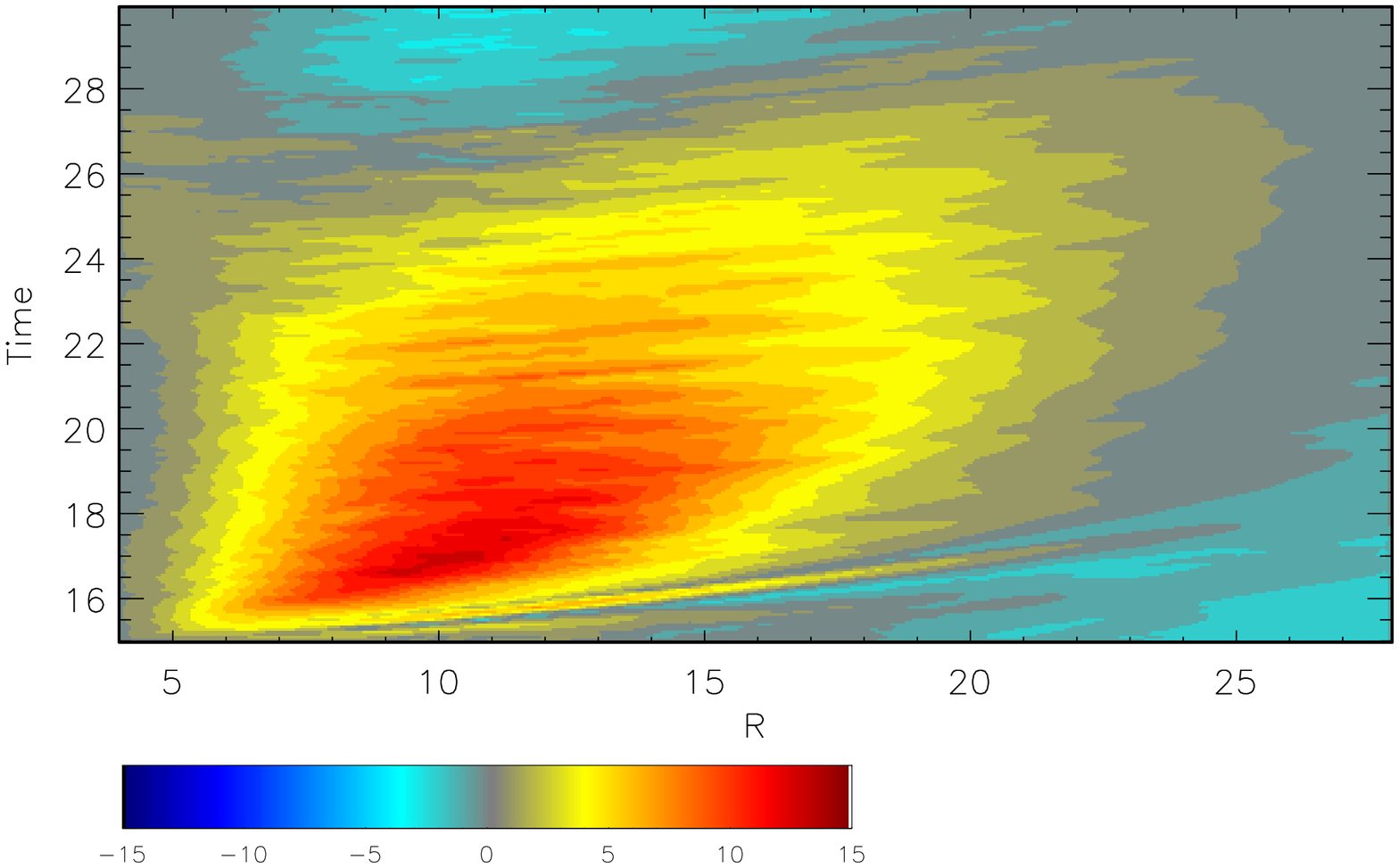} \\
\includegraphics[width=0.6\textwidth,angle=90]{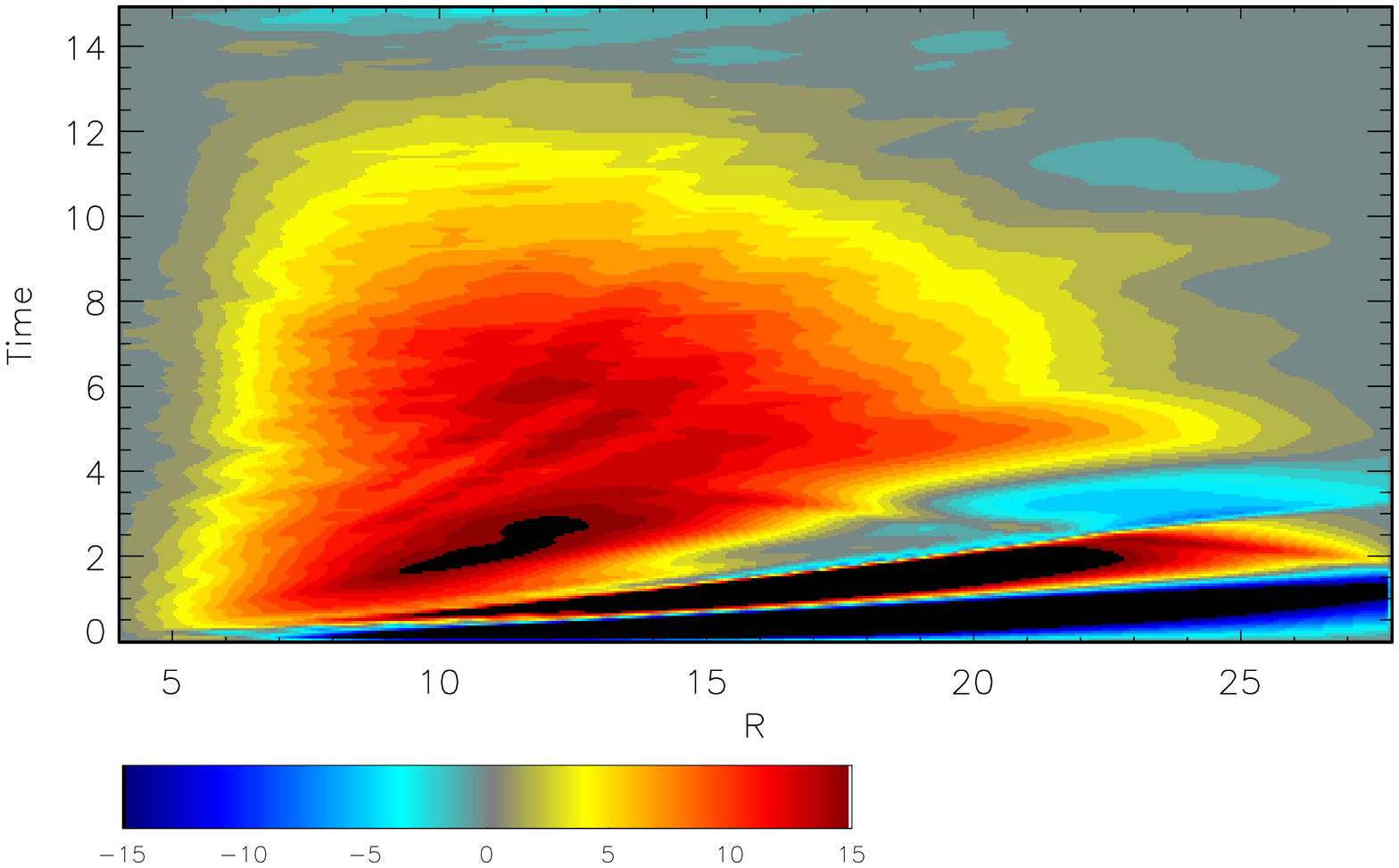} \\
\caption{Color contours (see color bar) of the radial flux of the $x$ component of
angular momentum.  Upper panel is \runm; lower panel is \runh .}
\label{fig:srx}
\end{center}
\end{figure}

\begin{figure}
\begin{center}
\includegraphics[width=0.6\textwidth,angle=90]{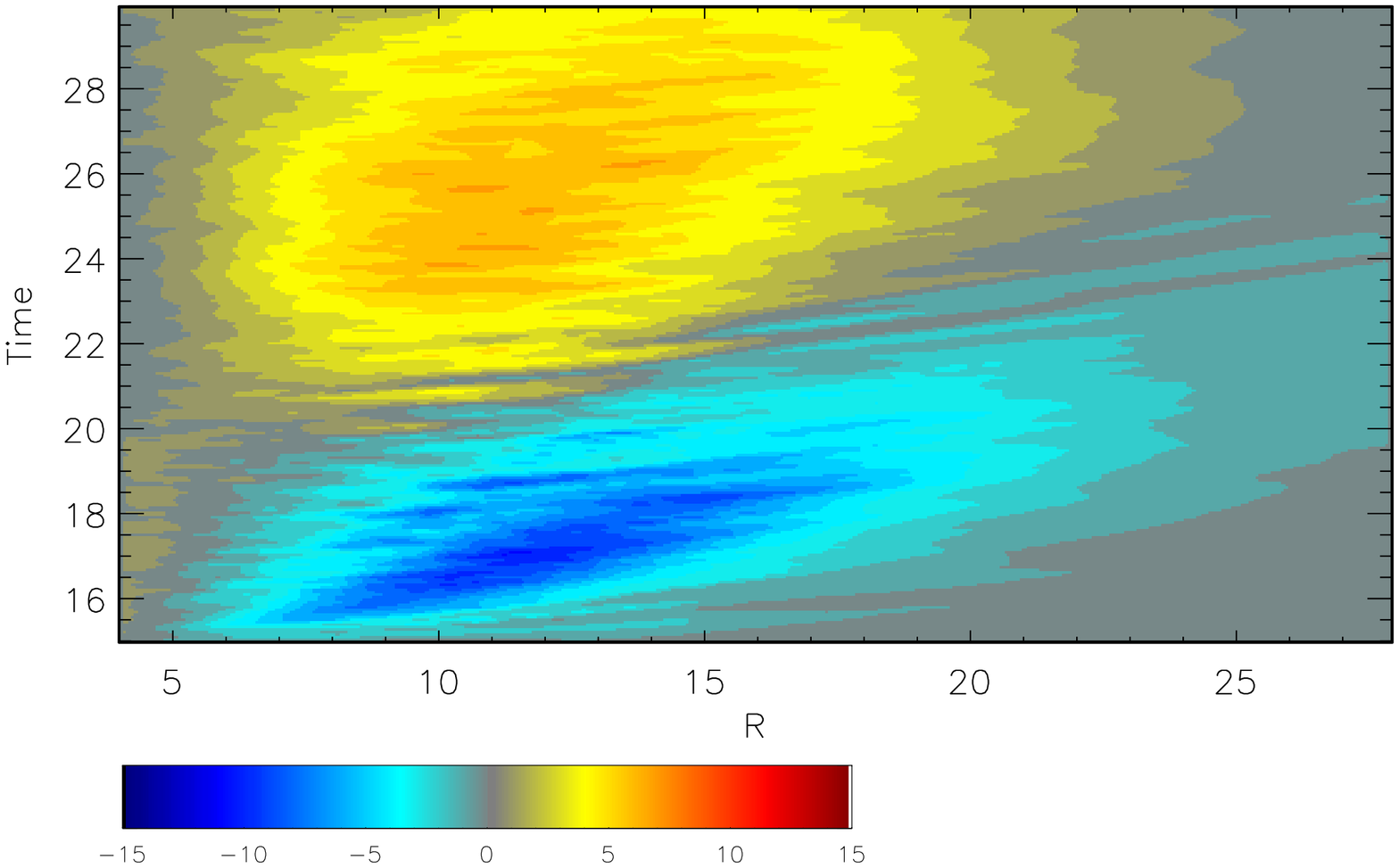} \\
\includegraphics[width=0.6\textwidth,angle=90]{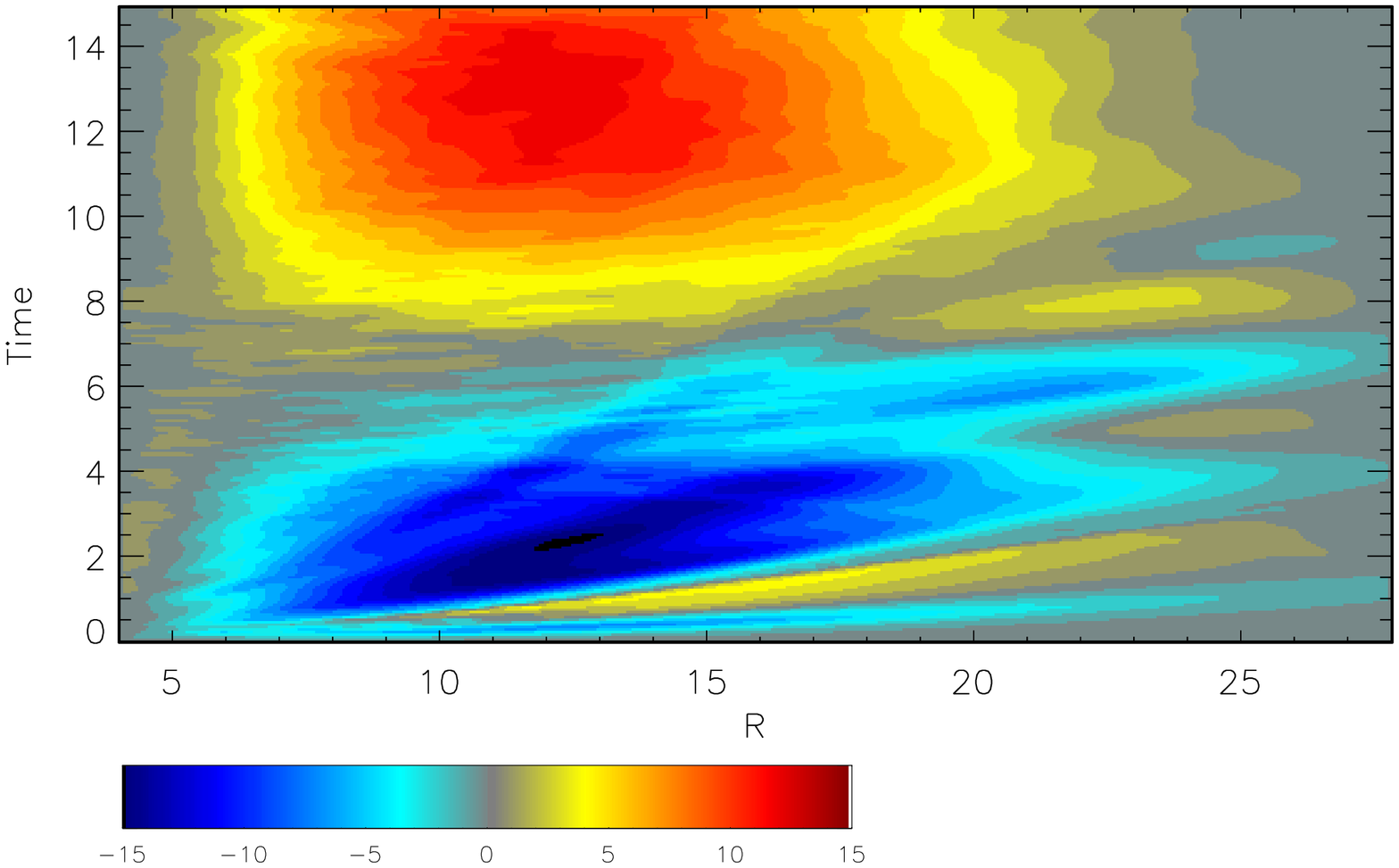} \\
\caption{Color contours (see color bar) of the radial flux of the $y$ component of
angular momentum.  Upper panel is \runm; lower panel is \runh .}
\label{fig:sry}
\end{center}
\end{figure}

More formally, we define the radial angular momentum flux as
\begin{equation}
S_{r,(x,y,\perp)} \equiv r^2 \int \, d\theta \sin\theta \, \int \, d\phi \, \rho v_r \ell_{(x,y,\perp)},
\end{equation}
where $\ell_{x,y,\perp}$ is the local specific angular momentum in the $x$, $y$, or
perpendicular (i.e., combining $x$ and $y$) direction.  The magnitudes of these fluxes
are shown in Figures~\ref{fig:srx} and \ref{fig:sry}.  The global shape of the
radius and time dependence of the fluxes is similar in the two simulations, and
in units of shell-integrated $\rho r v_{\rm orb}^2$, the magnitudes of the
fluxes in both simulations are similar to those seen in \cite{SKH13} when $\phat \simeq 1$.
In that previous paper, fluxes of this magnitude led to approximate disk flattening
on orbital timescales; much the same result is seen in both of these new simulations.

However, there are also significant contrasts.  Most importantly, in the hydrodynamic case,
but not in the MHD simulation, there is a sequence of three large amplitude flux pulses of
alternating sign.  The first two are due to a transient in which the hydrodynamic
disk relaxes from its initial state, which is not exactly in equilibrium; their effects
nearly cancel.  The third, although smaller in magnitude, is in the long-run more
significant.  It follows the track already seen in Figure~\ref{fig:psihat} that we
interpreted as a bending wave.  This track also corresponds to the flattening of the
disk seen in Figure~\ref{fig:phi_prec} and coincides with the last stage of alignment
seen in Figure~\ref{fig:align}.  In other words, it appears that this bending wave
pulse effectively flattens the hydrodynamic disk, so that it precesses very nearly
as a solid-body thereafter.  Once the bending wave has passed, \runh\ maintains
a generally higher level of outward angular momentum flux than found in \runm\ because
it remains misaligned to a greater degree at small radius where the torques operate
(see also Figs.~\ref{fig:Gx} and \ref{fig:Gy}).  By contrast, in \runm, although there
is an initial bending wave, it is partially disrupted, and is much less effective
at flattening the disk, as demonstrated also by the generally higher levels of
$\phat$ seen in the MHD panel of Figure~\ref{fig:psihat} between the tracks of the bending wave
and the kinematic precession pulse.

\begin{figure}
\begin{center}
\includegraphics[width=0.6\textwidth,angle=90]{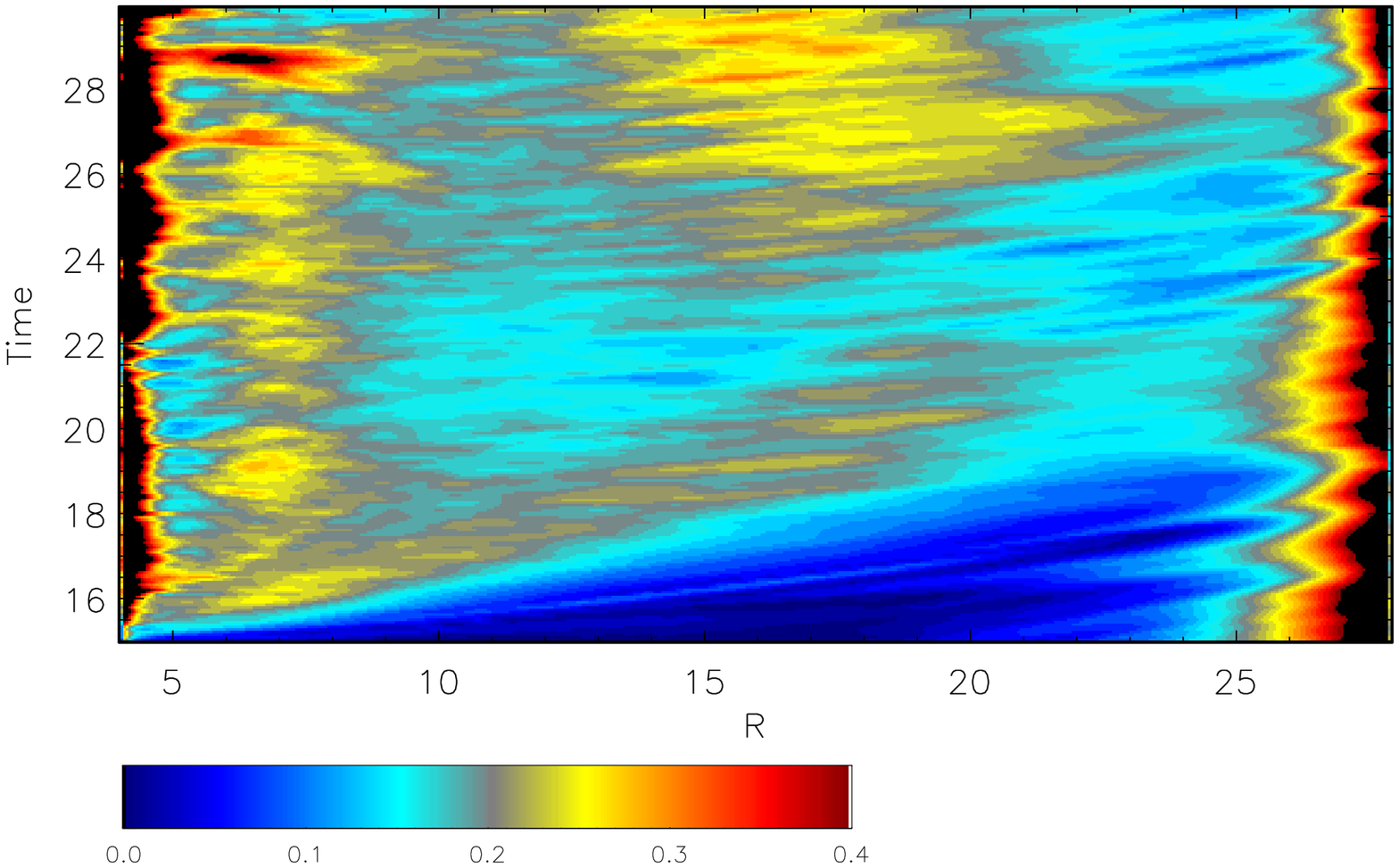} \\
\includegraphics[width=0.6\textwidth,angle=90]{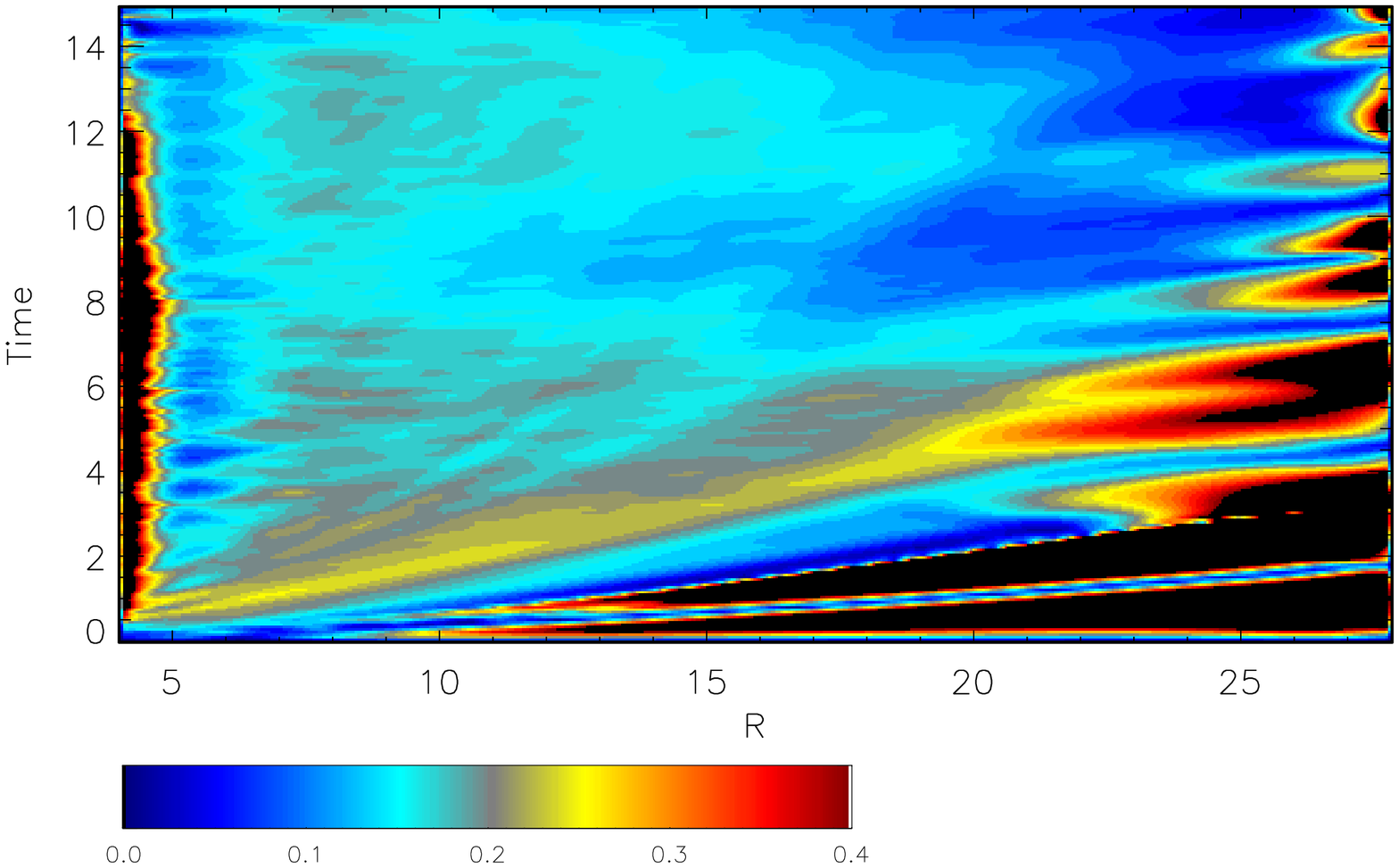} \\
\caption{Color contours (see color bar) of $\hat S_{r\perp}$, the normalized radial
flux of the perpendicular component of angular momentum.  Upper panel is \runm;
lower panel is \runh .}
\label{fig:srpnorm}
\end{center}
\end{figure}

To gain a sense of scale, it is also useful to look at a normalized version of
the angular momentum flux,
\begin{equation}
\hat S_{r\perp} \equiv \frac{|S_{r\perp}|}{c_s \partial L_\perp/dr}.
\end{equation}
This quantity captures the efficiency with which local fluid is able to pass along
its angular momentum.  Shown in Figure~\ref{fig:srpnorm}, we see that this quantity
is typically a few tenths; that is, if the mean flow rate is exactly the sound speed,
the flux carries $\sim 15$--30\% of the local angular momentum.  Although the
absolute level of the fluxes in the MHD case was always somewhat smaller than in
the HD case, $\hat S_{r\perp}$ is always larger in MHD.  In other words, the MHD
case puts more of its available misaligned angular momentum into motion.  The
contrast is especially noticeable in locations where the swing into alignment
is most rapidly taking place.
On this basis, it might be reasonable to identify
the magnitude of $\hat S_{r\perp}$ found here with $h/\Delta r$, where $\Delta r$
is the radial scale of the warp.   We caution, however, that, as shown by
\cite{SKH13}, the actual functional relationship between $S_{r\perp}$ and
the warp magnitude $\hat\psi$ is nonlinear, exhibits time delays, and also
depends on the global character of the warp.
Consistent with those results, $\hat S_{r\perp}$ varies by a factor $\sim 2$,
both as a function of time and as a function of radius.   Because $\hat S_{r\perp}$
is proportional to the ratio of angular momentum flux to the radial gradient of angular
momentum direction, these fluctuations support the conclusion of our previous
paper that simple diffusion model does not fully describe the behavior of this system.

\begin{figure}
\begin{center}
\includegraphics[width=0.6\textwidth,angle=90]{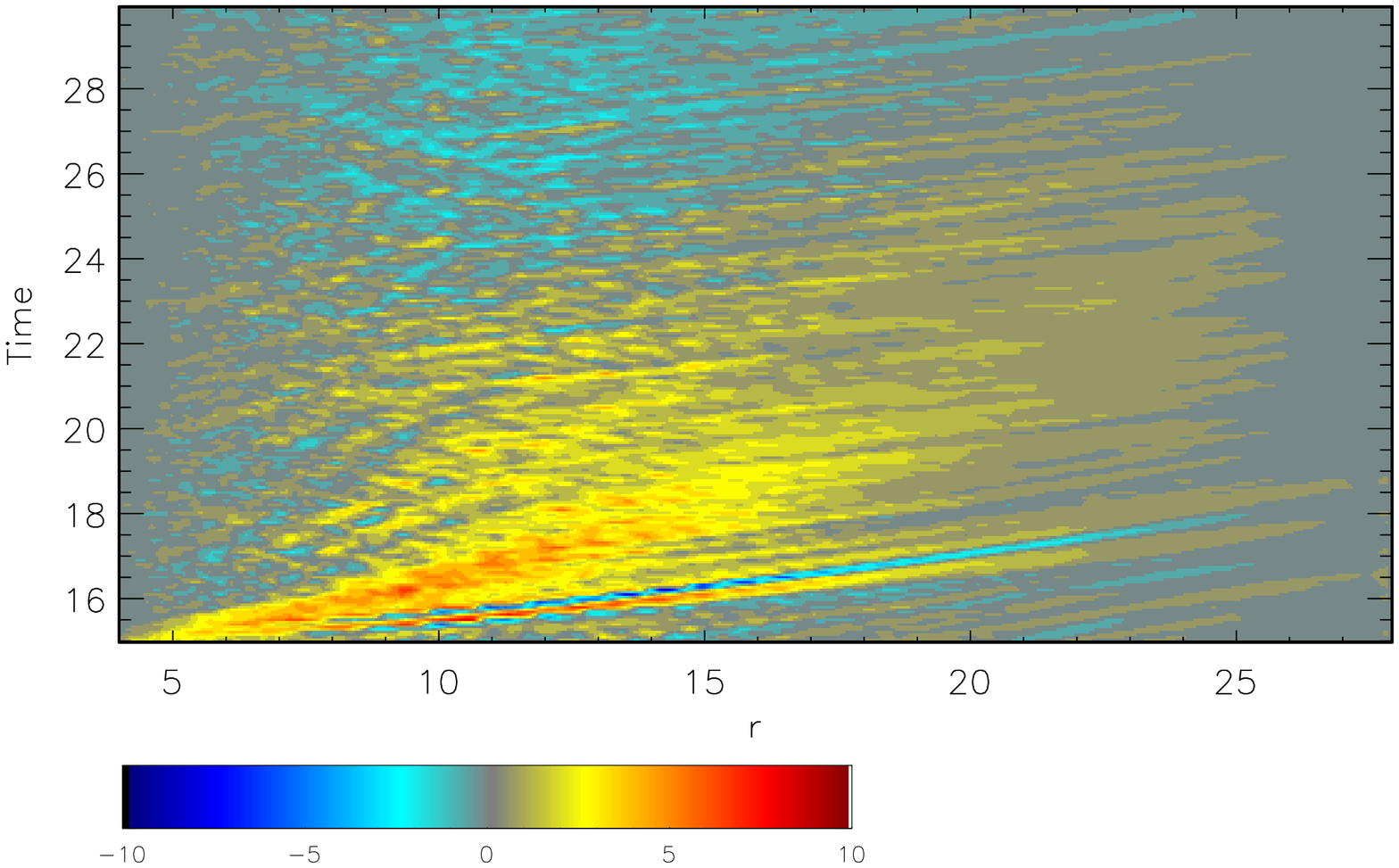} \\
\includegraphics[width=0.6\textwidth,angle=90]{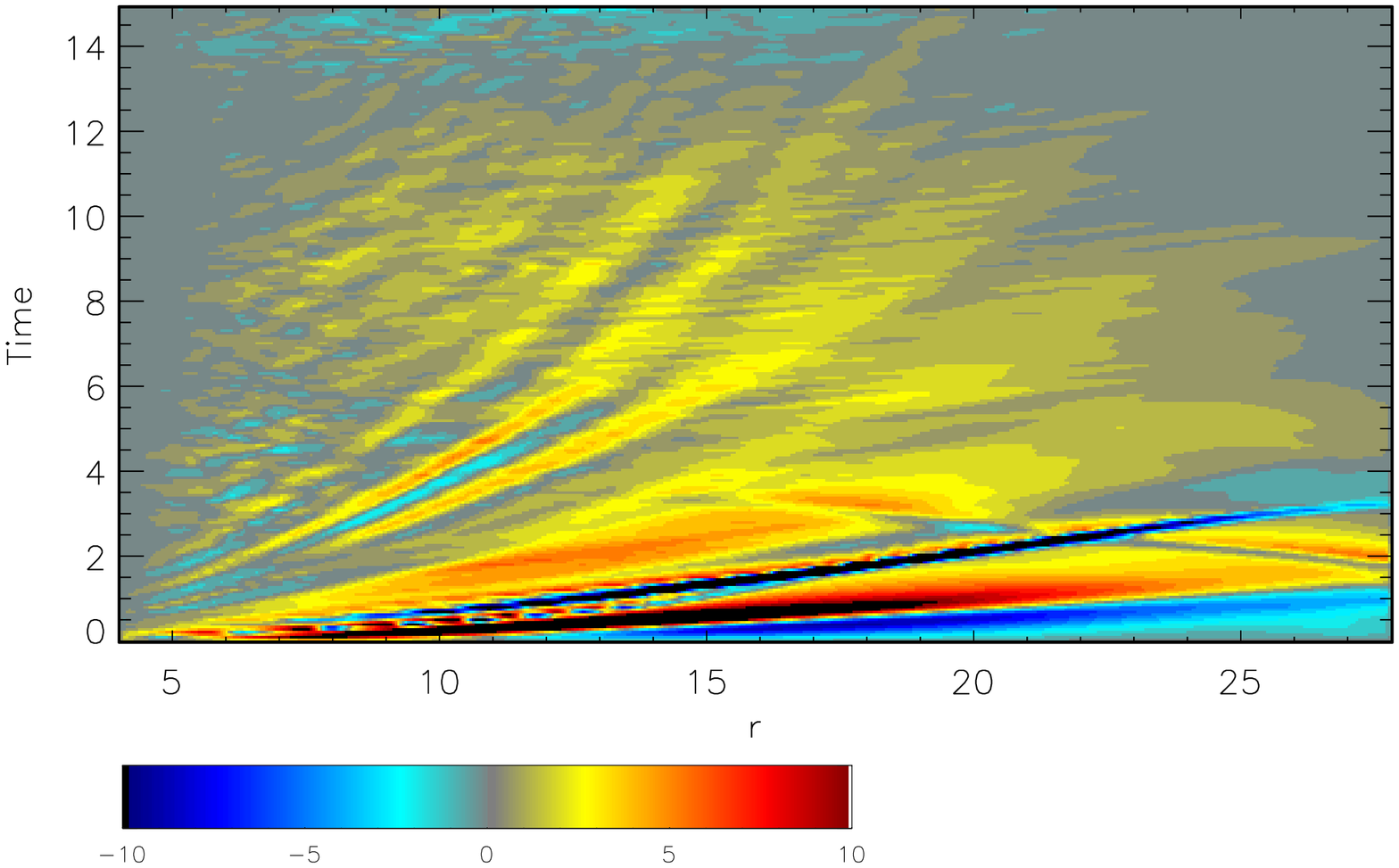} \\
\caption{Color contours (see color bar) of the net rate of change per unit
time in the $x$ component of angular momentum in each radial shell.
Upper panel is \runm; lower panel is \runh .}
\label{fig:netx}
\end{center}
\end{figure}

\begin{figure}
\begin{center}
\includegraphics[width=0.6\textwidth,angle=90]{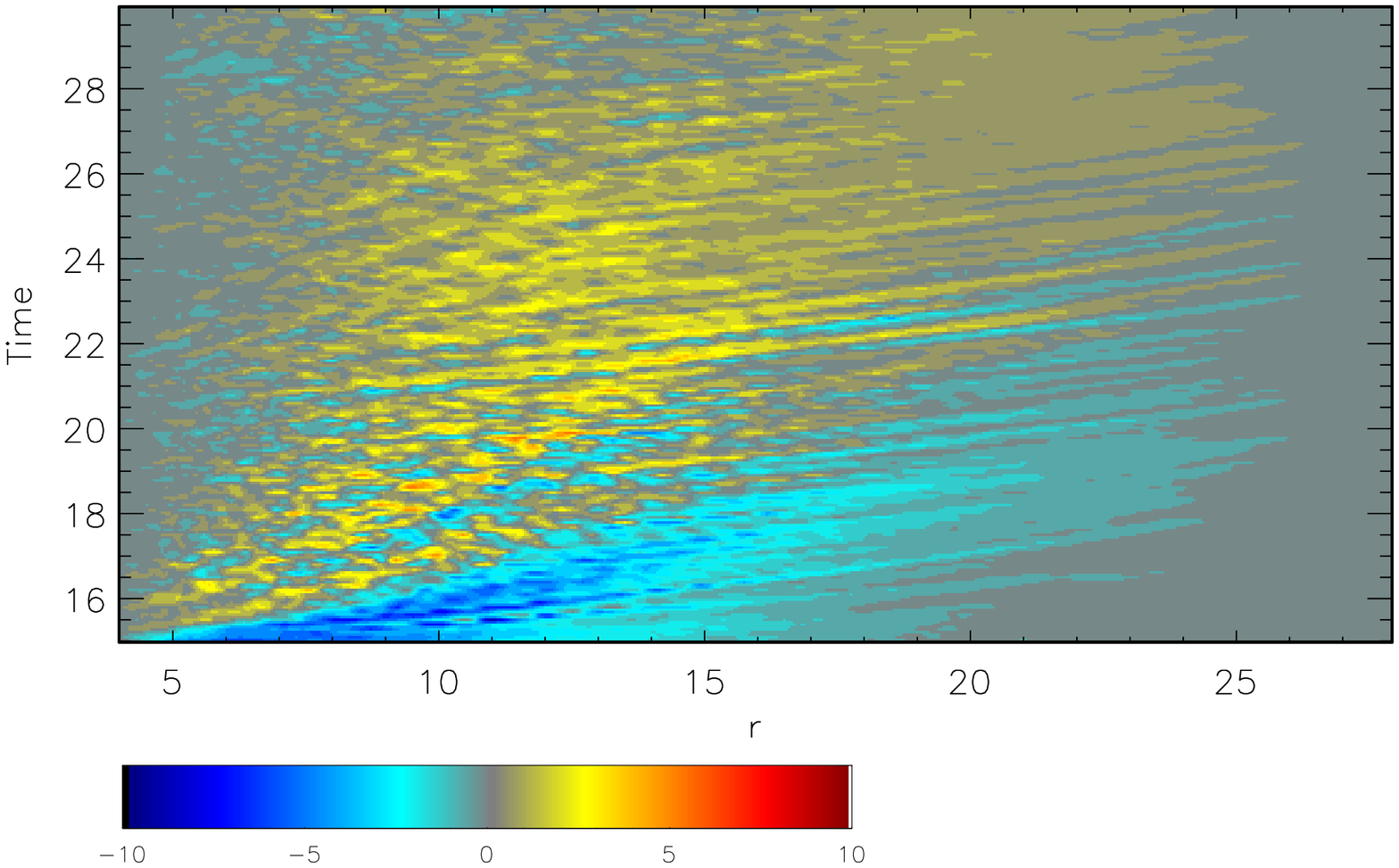} \\
\includegraphics[width=0.6\textwidth,angle=90]{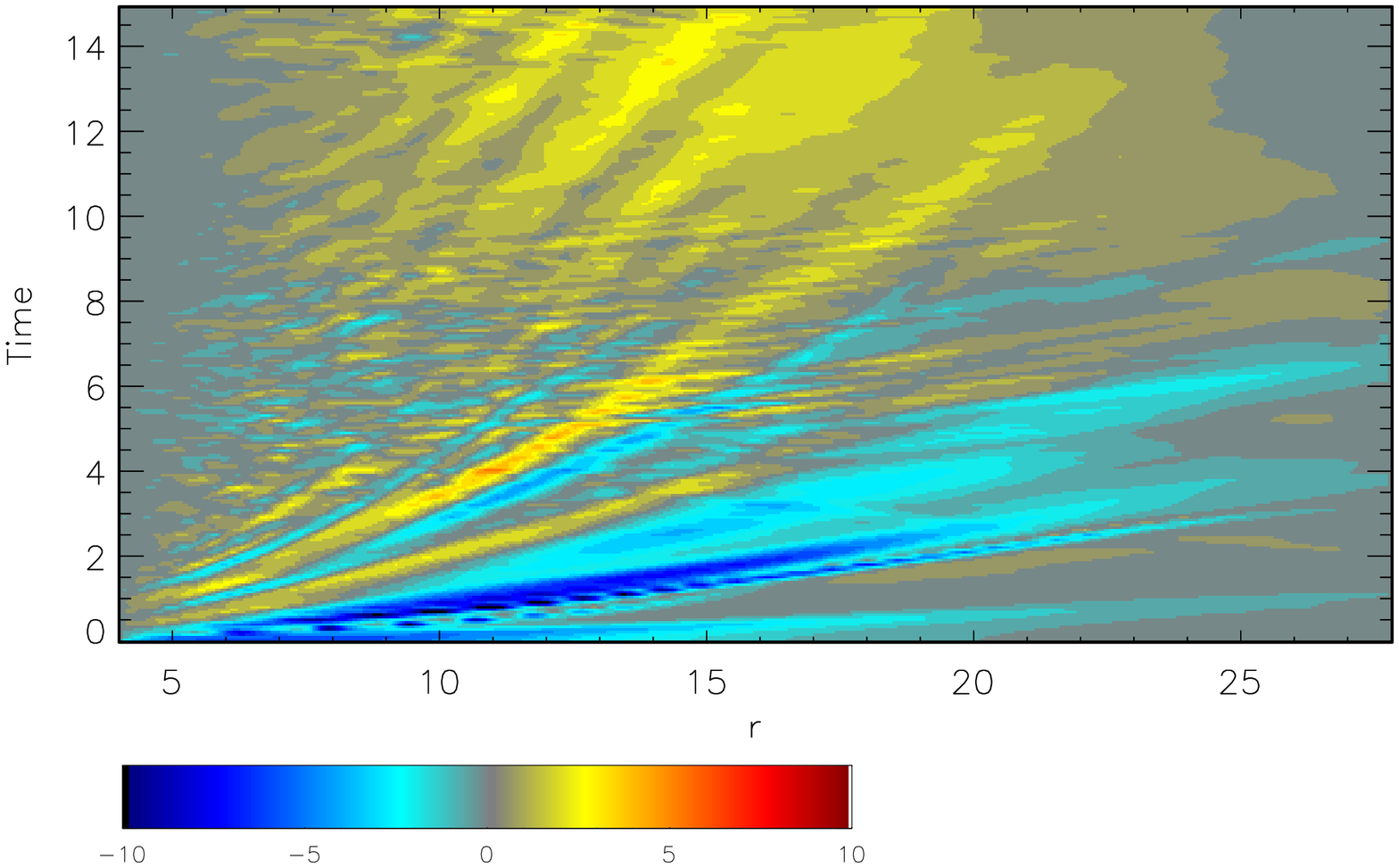} \\
\caption{Color contours (see color bar) of the net rate of change per unit
time in the $y$ component of angular momentum in each radial shell.
Upper panel is \runm; lower panel is \runh .}
\label{fig:nety}
\end{center}
\end{figure}

The next step in our inquiry is to examine the effect of divergence in
the angular momentum flux.  Comparing Figures~\ref{fig:srx} and \ref{fig:sry} with
their torque counterparts (Figs.~\ref{fig:Gx} and \ref{fig:Gy}, respectively), it is
apparent that the fluxes are largest at radii considerably greater than where the
torques are largest.  In other words, the angular momentum delivered at small radii
by the torques is collected and swept outward.  Beyond $r \simeq 14$--17, where the
fluxes peak, the transported angular momentum is deposited, a bit like silt dropping out of
a slowing river.  The distribution of the net rate of change in angular momentum
can be seen in Figures~\ref{fig:netx} and \ref{fig:nety}.  Both the initial bending wave
and the later, slower pulses seen in Figure~\ref{fig:psihat} can be discerned in
the HD panel of Figure~\ref{fig:netx}.  These pulse trains are much less apparent in the MHD
case.
One fact uniting the
HD and MHD plots of Figures~\ref{fig:netx} and \ref{fig:nety}, however, is that
in both cases the net rate of change in angular momentum in regions where the
rate of angular momentum delivery by torque is high is in fact quite small.
In other words, where the torque is delivered, increased outward angular momentum
flux removes the great majority of it and transports it outward.  This is
why the local precession rate in the inner disk is substantially smaller
than the test-particle model would predict and also why, especially in \runm,
much of that angular momentum is used for alignment rather than precession.

The effects shown in Figures~\ref{fig:netx} and \ref{fig:nety} can be summarized
by the fact that
\begin{equation}\label{eqn:alignmentrate}
\frac{\partial^2 L_\perp}{\partial r\partial t} = 
   -\frac{\partial L_x}{\partial r}\frac{\partial S_{rx}}{\partial r} -
                         \frac{\partial L_y}{\partial r}\frac{\partial S_{ry}}{\partial r}.
\end{equation}
If the ratio $(\partial S_{rx}/\partial r)/(\partial S_{ry}/\partial r)$ were equal
to $G_x/G_y$, the perpendicular angular momentum would not change at all because
that is the condition for precession.  However, the ratio of the angular momentum
deposition rates in the $x$ and $y$ directions does not necessarily match the ratio
required for precession at that location.  To accomplish alignment, all that is
required is for
\begin{equation}
\frac{\partial S_{sry}/\partial r}{\partial S_{rx}/\partial r} < 
    -\frac{\partial L_x/\partial r}{\partial L_y/\partial r},
\end{equation}
where the RHS is the exact precession ratio.  Alignment proceeds most rapidly where
this inequality is most strongly satisfied.

Another way of putting the same point is to observe that alignment is
achieved most efficiently when the vector
\begin{equation}
\frac{\partial^2 {\bf L_\perp}}{\partial r\partial t} =
    -\frac{\partial S_{rx}}{\partial r}\hat x
                         - \frac{\partial S_{ry}}{\partial r}\hat y
\end{equation}
is exactly anti-parallel to $\partial {\bf L_\perp}/\partial r$.  In principle
the angle $\gamma$ between $-\partial {\bf L_\perp}/\partial r$ and the rate
at which it is changed might be anywhere from 0 to $\pi$.
The angle optimally efficient for alignment is $0$; the angle that
produces pure precession is $\pi/2$.
In both \runh\ and \runm, we find that during times of alignment
$\langle\cos\gamma\rangle \simeq 0.5$ although there are sizable fluctuations
around this value at specific times and locations.  On the other hand,
$\langle\cos\gamma\rangle$ decreases over time in \runh\ from
approximately this value during the first $\simeq 6$~orbits to close to zero during
the remainder of the simulation.

In large part, the angle $\gamma$ is determined by the relative precession
angles of the region where the torque occurs, which supplies the angular
momentum for the outward flux, and the region where the angular momentum
is deposited.  For maximal alignment rate, the direction of the deposited
angular momentum should be the direction of the torque exerted when the precession
angle is $\pi/2$ in advance of the local precession angle.  The value of
$\langle\cos\gamma\rangle$ seen in \runm\ indicates a precession angle difference
closer to $\simeq \pi/6$ than $\pi/2$, but there is nonetheless sufficient offset to
drive alignment.  During early times in \runh\, the situation is similar.
At late times in \runh\, however, $\langle\cos\gamma\rangle \simeq 0$ because
the disk orientation is very nearly the same at all radii, and the time
required to transport angular momentum from the small radii where the
torques operate to larger radii is short compared to the solid-body
precession period.  In other words, having at least some warp in the
disk is essential to alignment.

\begin{figure}
\begin{center}
\includegraphics[width=0.6\textwidth,angle=90]{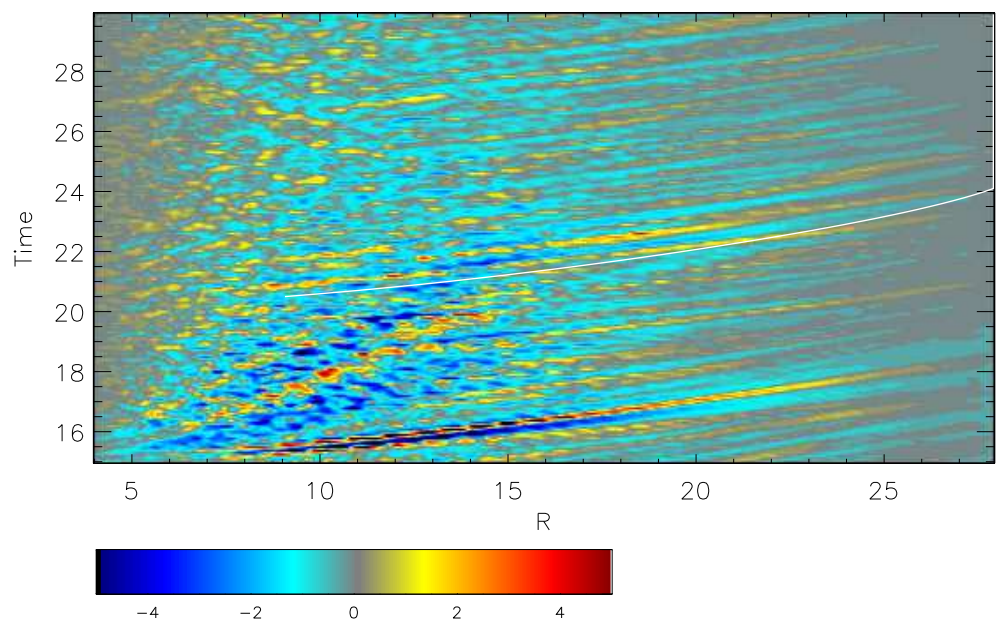} \\
\caption{Color contours (see color bar) of $\partial \ln L_\perp/\partial t$ in
\runm.  The white curve shows the trajectory of a sound wave traveling radially
outward.}
\label{fig:alignmentrate}
\end{center}
\end{figure}

The detailed radial and time dependence of the net rate of change of misaligned
angular momentum $\partial^2 L_\perp/\partial r\partial t$ in \runm\ is shown in
Figure~\ref{fig:alignmentrate}.  Several things
stand out in this plot.  One is that the local rate of change of misaligned angular
momentum is predominantly, but not exclusively, negative.  That is,
there are frequently moments when an individual ring becomes {\it less},
not {\it more} aligned, even though the long-term trend is toward alignment.
Another point is that, not surprisingly, the largest
part of the change in angular momentum is associated with the range of radii
($8 \lesssim r \lesssim 15$) with the greatest mass and therefore the greatest
amount of misaligned angular momentum to change.

Perhaps more surprisingly, this figure is also marked by a large number of
streaks indicating rapid outward motion.  The white curve in the
figure follows the path of an adiabatic sound wave directed radially outward.
The very close correspondence between its slope in this diagram and the slopes
of the streaks demonstrates clearly that these are the traces of sound waves.
Although they are certainly not regular, there is a typical time interval
between these waves, $\simeq 0.5$ fiducial orbits.

\begin{figure}
\begin{center}
\includegraphics[width=0.6\textwidth,angle=90]{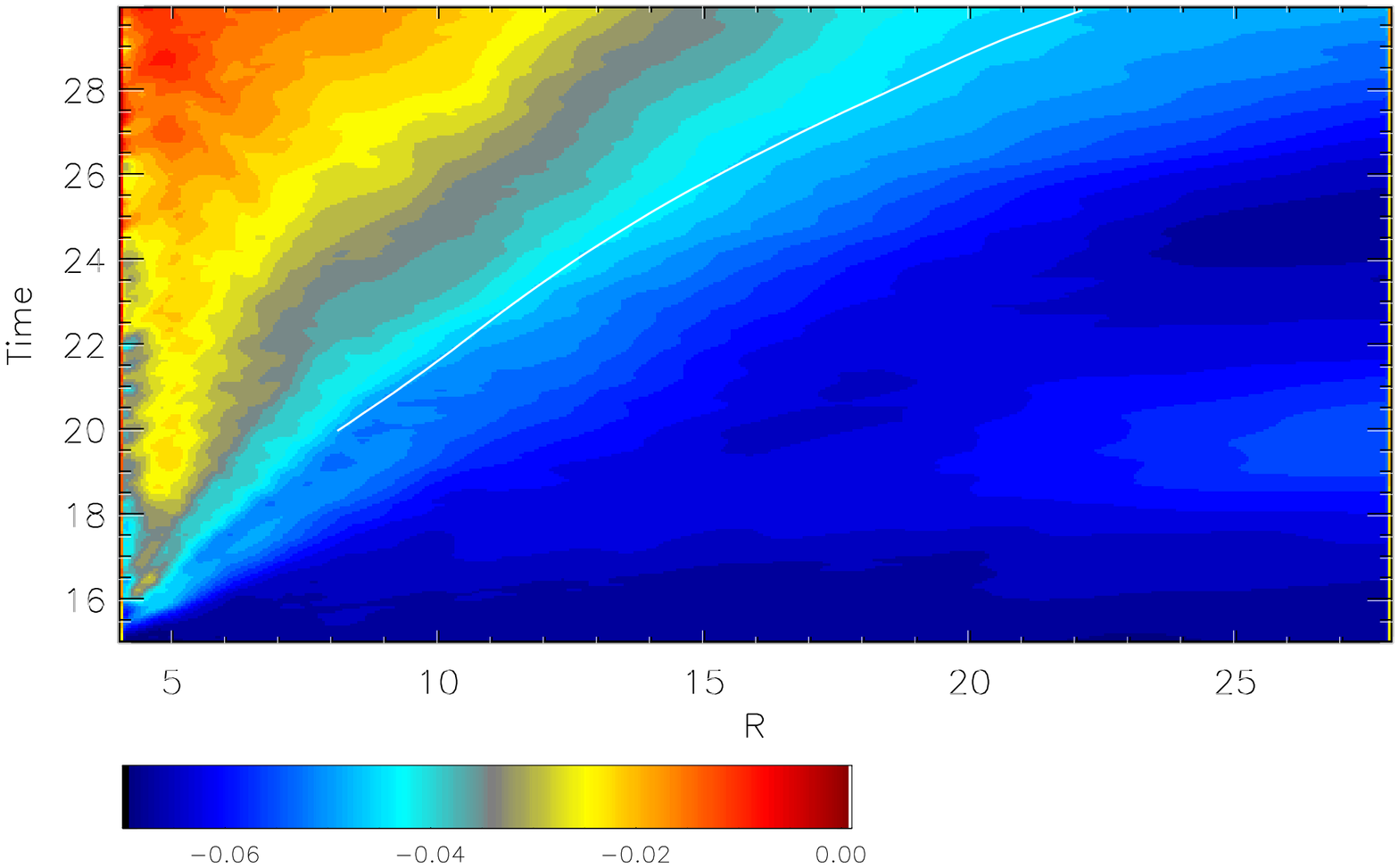} \\
\caption{Color contours (see color bar) of the inclination angle (in units of $\pi$)
in \runm.  The white curve shows the path of the alignment front traveling at the speed
indicated by Eqn.~\ref{eqn:alignrate}.}
\label{fig:alignadj}
\end{center}
\end{figure}

In \runm, alignment is achieved beginning at small radii.  Consequently, one
can speak of the outward motion of an alignment front.  On dimensional grounds,
one might estimate the rate at which this front moves outward by the ratio of
the rate at which unaligned angular momentum is given to the disk by torque
to the magnitude of the local angular momentum requiring alignment.
However, as the previous discussion would suggest, this estimate should be corrected
by a factor $\langle \cos\gamma\rangle$.  Our estimated rate of motion would
then be
\begin{equation}\label{eqn:alignrate}
\frac{dr_f}{dt} = \langle\cos\gamma\rangle \frac{G(<r_f)}{dL_{\perp}(r_f)/dr},
\end{equation}
where $r_f$ is the radius of the alignment front and $G(<r)$ is the magnitude of the torque
integrated over the matter interior to $r$.
At the order of magnitude level, $dr_f /dt \sim \Delta r_f \omega$, where
$\Delta r_f$, the radial width of the alignment front, is $\sim r_f$ in \runm.
This estimate might also be further reduced by an allowance for some of the
torque being deposited at radii between where it is given to the disk and
the alignment front.  However, if the transition region from an aligned inner
disk to the inclined outer disk is reasonably narrow, this loss may not be
very large.  Confirmation of this guess is provided in
Figure~\ref{fig:alignadj}, where we show the track of an alignment front
moving at the speed we estimate assuming $\langle \cos\gamma\rangle = 0.5$.
As can be seen, it follows the contour of half-alignment quite well.

This model as stated implicitly assumes that the angular momentum given the
disk by the external torques is instantaneously mixed radially across the entire
transition region.   In reality, of course, the mixing speed is of order the radial
flow speed, which, we have argued, is roughly the sound speed in the presence
of nonlinear warps.   However, as shown by \cite{SKH13}, the Mach number of these radial
motions is quite sensitive to $\hat\psi$ and the radial width of the warp, so that
the sound speed is at best a rather crude estimator of the mixing speed.   Nonetheless,
in the conditions of \runm, the sound speed is $3$--$6 \times$ larger than the
alignment front speed, so that our instantaneous delivery approximation has little
effect.    A slower sound speed might lead to a narrower transition region.

\subsection{Alignment, stalled and completed}

As is readily apparent in Figure~\ref{fig:align}, although \runh\ diminishes its
misalignment, it is never able to remove more than $\simeq 40\%$ of its tilt,
whereas \runm\ continuously eliminates the offset between its angular momentum
direction and the central mass's spin, achieving hardly any difference between
the two throughout its inner radii by the end of the run.  Given that even in
\runm\ magnetic forces are thoroughly dominated by pressure forces, what accounts
for this contrast?

We suggest that the answer lies in a combination of two facts.  First, as we
have already mentioned, the HD case rapidly achieves a state in which it precesses
nearly as a solid-body.  In the MHD case, by contrast, turbulence
interferes with the ability to enforce solid-body rotation.  As a consequence,
in \runh\ but not \runm\, the
direction of the planar angular momentum brought to a given radius is close to
the direction of precession torque.  In the language of the preceding section,
after the first $\sim 5$~orbits or so, $\cos\gamma \simeq 0$ in \runh.

Second, as found by \cite{Sorathia12}, the Reynolds stresses capable of mixing angular
momentum radially are strongly increasing functions of disk warp when $\hat\psi > 1$.
Below that level of warp, the radial pressure gradients are incapable of driving
radial motions to speeds comparable to or greater than the sound speed; above
that level, such speeds are generically attained.  Comparing Figures~\ref{fig:align}
and \ref{fig:psihat}, one can see that alignment drastically slows in the HD
case when $\hat\psi$ drops below unity, in line with the expectation that
when the tilt angle becomes $< H/r$, the warp-induced Reynolds stresses weaken.
Because a purely hydrodynamic disk is always laminar, the alignment process therefore
stops at this point.  The MHD case differs because the magneto-rotational instability
insures ubiquitous turbulence.  Even where $\hat\psi$ is too small to drive strong
radial flows, MHD turbulence nonetheless continues to mix neighboring regions.
It is this process that allows MHD turbulence to complete the work of alignment
after Reynolds stresses reduce the misalignment angle to be only of order the
disk aspect ratio.

\subsection{The inclination transition radius in an accreting disk}

We have already estimated the alignment speed in terms of the torque integrated
interior to some radius relative to the unaligned angular momentum at that radius.
Presumably, if we had run simulation \runm\ still longer, the alignment front
would have propagated all the way out through our finite disk, and that would
have been the end of further evolution.  In real disks, however,
the reservoir of matter with inclined angular momentum extends much farther out,
and new unaligned matter is continually fed from the outside, while matter already in the disk
gradually moves inward toward the central object.  Because the alignment
speed inevitably must diminish outward as the torques weaken, in such
a disk an outwardly moving alignment front would eventually find itself moving
so slowly relative to the inward flow of
misaligned angular momentum
that its motion relative to the central mass would be reduced to zero.  Thus, in a disk
with time-steady accretion, both in terms of mass inflow and orientation,
and a central object with a mass very large compared to the accreted mass,
the disk would bend from its initial orientation to the orientation of the central
object's angular momentum at a fixed transition radius where these two speeds cancel.

The local torque scales with the surface density and $\sin\beta$, for misalignment
angle $\beta$.  The local misaligned angular momentum does likewise.  Because
the alignment front propagation speed is proportional to the ratio between
the integrated torque interior to a given radius and the misaligned
local angular momentum, it is therefore
\begin{equation}
\frac{d r_f}{dt} = \frac{2 \langle \cos\gamma \rangle a_* (GM)^2}{\sin\beta(r) c^3 r^{3/2} \Sigma(r)}
 \int_0^r \, dr^\prime \sin\beta(r^\prime)\Sigma(r^\prime)/{r^\prime}^{3/2}
\end{equation}
for black hole spin parameter $a_* \equiv a/M$.

In real disks, fresh misaligned angular momentum can be brought inward either by
accretion of new material or by warp-induced radial flows; gravitational interaction
with a binary companion or the mass of the outer disk
may also contribute \citep{Tremaine2013}.    Outward motion
of the alignment in mass terms can then be brought to a halt in terms of position
when that inward speed matches the outward progress of the alignment front.
Parameterizing the characteristic timescale of the inward advection of misaligned
angular momentum by $t_{\rm in}$, we find that the time-steady position of the inclination
transition can be estimated as
\begin{equation}
R_T/r_g = \left[ 2\langle \cos\gamma\rangle a_* \Omega(R_T) t_{\rm in}
   \int_0^1 \, dx \, x^{-3/2} \frac{\sin\beta(x)}{\sin\beta(R_T)}
   \frac{\Sigma(x)}{\Sigma(R_T)} \right]^{2/3},
\end{equation}
where $r_g \equiv GM/c^2$ and the integral has been nondimensionalized by
setting $x = r^\prime/r = r^\prime/R_T$.

The dimensionless integral may often have a value rather greater than unity. 
Partly this is due to an effect we have already pointed out, that the rapid increase
inward of the precession frequency allows inner radii that are already nearly
aligned to account for a significant part of the total torque.   In addition, however,
in some commonly-encountered accretion regimes, the surface density also
increases inward (in time-steady accretion, for example, $\Sigma \propto x^{-3/5}$
when gas pressure dominates radiation pressure and the principal opacity is
electron scattering).    The dimensionless integral would then be rather greater
than unity when the outermost part of the transition region lies at a radius a factor
of a few or more greater than the innermost part.   
This happens, for example, in the later stages of \runm\ when $r_f$ (as defined
by the white curve in Fig.~\ref{fig:alignadj}) passes
the radius of maximum surface density, $r \simeq 10$.   From then onward,
the dimensionless integral is $ > 1$, reaching $\simeq 5$ or more by the end of the
simulation, when $r_f \simeq 22$.   However, we caution that this is at best
illustrative: the detailed shape of the alignment transition is
likely to be influenced by a number of factors: in addition to the shape of the radial surface density
profile, the disk thickness profile, and perhaps other variables may also matter.

If the dominant misaligned angular momentum inflow mechanism is accretion, the inflow speed
$v_{\rm in} \simeq \alpha (h/r)^2 v_{\rm orb}$,
where $\alpha$ is the usual ratio between integrated internal (Maxwell) stress
and integrated pressure, $h/r$ is the local aspect ratio of the disk, and
$v_{\rm orb}$ is the Keplerian orbital velocity.   In this case, we find
\begin{equation}\label{eqn:RTinflow}
R_T/r_g = \left[ \frac{2\langle \cos\gamma\rangle a_*}{\alpha (h/R_T)^2}
   \int_0^1 \, dx \, x^{-3/2} \frac{\sin\beta(x)}{\sin\beta(R_T)}
   \frac{\Sigma(x)}{\Sigma(R_T)} \right]^{2/3}.
\end{equation}
At the order of magnitude level, this estimate is consistent with the original
estimate given by \cite{BP75} and \cite{Hatchett81}, although there are also
ways in which our estimate differs from theirs.  In particular, we note the
quantitative importance of the dimensionless integral in equation~\ref{eqn:RTinflow}.

Not long after these original estimates, \cite{PP83} argued that the radial flows
driven by warping should carry misaligned angular momentum much more rapidly than
the mass flow of accretion.   Moreover, in the model presented by that
paper and elaborated by many since \citep{Pringle92,NP00,LP07}, the inward
mixing can be described as a diffusion process with effective diffusion
coefficient $\alpha_2 \simeq 1/(2\alpha)$ when $\alpha \ll h/r$.   In that
case, the estimate for $R_T/r_g$ is (modulo the dimensionless integral)
identical to that of eqn.~\ref{eqn:RTinflow}, but multiplied by $\alpha^{4/3}$.

However, our previous study of purely hydrodynamic warp relaxation \citep{SKH13}
demonstrated that, while the radial mixing of misaligned angular momentum
qualitatively resembles diffusion, it differs from diffusion in a number of
quantitative aspects;
even in SPH simulations with an isotropic viscosity,
the diffusion approximation appears to break down for nonlinear warps \citep{Lodato10}.
In addition, as shown in Section~\ref{sec:mhdvshd} of
this paper, there are no stresses limiting the radial motions in a fashion described
by an isotropic ``$\alpha$-viscosity"; consequently, there is no reason to expect
the radial mixing to scale $\propto \alpha^{-1}$.

A better estimate of the inward mixing rate might be
$\sim c_s^2/v_{\rm orb}(r_f/\Delta r_f)^2$,
similar to the speed \cite{NP00} identify with the case in which $\alpha \sim h/r$.
This estimate is also closer to the rate found by \cite{LP07} and \cite{Lodato10}
when $\alpha$ was small; in that limit, their SPH simulations with an isotropic $\alpha$ viscosity
indicated that $\alpha_2$ saturated at $\simeq 3$.
The basis of our estimate is that,
as found in Sec.~\ref{sec:align}, $\hat S_{r\perp} \sim h/\Delta r_f$ because
nonlinear warps generically create transonic radial flows, and they can travel a
distance $\sim h$ in radius before being turned back by gravity.   It must be
recognized, however, that this is a very rough estimator, as it hides the fact that
the magnitude of the misaligned angular momentum flux also depends
on the shape of the transition region \citep{SKH13}.   The simulation
data presented here do not bear directly on the effectiveness of inward mixing
because the adherence of the
alignment front propagation to our model (as well as the detailed radial dependence
of the angular momentum flux) demonstrates that inward radial mixing plays at most
a minor role in \runm.  Thus, the best we can do here is place bounds on $R_T$: the
estimate of equation~\ref{eqn:RTinflow} is likely a solid upper bound, while a rough
lower bound is given by the same expression with $\alpha \sim 1$.

\section{Conclusions}
\label{sec:con}

We have carried out the first calculation of the Bardeen-Petterson effect in which
the internal stresses are grounded entirely in known physical mechanisms (i.e.,
Reynolds and Maxwell stresses), and the disk configuration is thin enough that warp
relaxation can be separated from accretion.
It is also the first calculation making use of physical internal stresses in which
disk alignment is observed.  As predicted early on by analytic arguments \citep{PP83},
the heart of the mechanism is the creation of radial pressure gradients due to
the warps induced by the radial gradient in the Lense-Thirring precession rate.
These radial pressure gradients drive radial fluid flows that convey misaligned angular
momentum with them.
By radially mixing misaligned angular momentum, these flows help to bind together
the disk, compelling it to rotate almost as a solid body.  In addition,
as these flows move outward through the disk, the direction of the misaligned
angular momentum they carry can, given some departure from solid-body precession,
become sufficiently opposed to the local direction
that, when mixed, the result is a reduction in the net magnitude of misaligned
angular momentum.  In this fashion, the disk gradually aligns with the spin axis
of the central mass, first in its inner portions and later at larger radii.

By contrasting a pair of matched simulations, one including MHD, the other including
only pure hydrodynamics, we were able to highlight the effects due to MHD
and clarify those depending only on hydrodynamics.
When the local warp is nonlinear (the generic situation), the radial flows are
always transonic in speed.  Internal stresses other than pressure
are much too small to significantly influence them; in particular, we find
no evidence for anything resembling an ``isotropic $\alpha$ viscosity"
acting to limit these radial motions.  Although the magnetic forces are always
small compared to pressure forces, they nonetheless have a significant effect
on both disk precession and the rate at which disks align with the angular
momentum of the central mass they orbit.  In particular, MHD effects cause
more rapid alignment and more complete alignment---hydrodynamic
alignment appears to stall when the offset angle falls to a value comparable
to the disk aspect ratio $H/r$.

We believe there are two reasons for this contrast, both due to the omnipresent
MHD turbulence.
The first is that MHD turbulence disrupts the phase coherence
of bending waves without necessarily damping them.   By doing so, it prevents
the enforcement of solid-body precession that occurs in the purely hydrodynamic case.
As a result, the angular momentum delivered at small radii has a component
directed {\it antiparallel} to the misaligned angular momentum at the
rather larger radii where that angular momentum is ultimately deposited
by the radial flows.  This is the central mechanism of alignment.
Outward carriage of ``corrective" angular momentum is essential because
the Lense-Thirring torques diminish so rapidly with increasing radius
that an inner radius with only a small remaining inclination may nonetheless
feel a greater torque than a radius only a factor of a few farther away
that has a considerably larger inclination.   The second effect of MHD
turbulence is that it continues to mix unaligned angular momentum, even when the local
warp is small enough (less than a scale height) that the radial motions
induced by pressure gradients due to the warp are weak.

Our detailed treatment of the disk's internal dynamics also reveals that
the flow of misaligned angular momentum within the disk is by no means
smooth and regular.  Radial fluxes of angular momentum are sharply increased
when the radial gradients in the Lense-Thirring torque build a local disk
warp whose angular contrast across a radius is at least as large as the disk
scale height.  The evolution of orientation at a fixed radius is therefore
a sort of ``stick and slip" process in which differential torque gradually
builds local warp, which is then erased quickly when it becomes nonlinear.
Radial propagation of angular momentum is further modulated by acoustic waves.


This picture suggests a model for the speed at which alignment in an initially-inclined
disk moves outward: the speed of the alignment front $v_f \simeq 0.5 G(<r)/(dL_\perp/dr)$, where the
factor 0.5 comes from computing the mean (anti-)alignment between the
angular momentum brought outward from the torqued regions and deposited at
the alignment front.   In a time-steady disk, the outward motion is eventually
brought to a halt by inflow of misaligned angular momentum due to a combination
of radial mixing induced by disk warp and the accretion flow itself.
To order of magnitude accuracy, we can use this picture to estimate where
the transition from inclined to aligned orbits takes place in a time-steady disk.
If all the uncertainties associated with the rate of inward misaligned
angular momentum flux and the radial distribution of torque are wrapped
into a single parameter $\Phi$, we suggest that its magnitude may be roughly
bounded by $1 < \Phi < \alpha^{-2/3}$, where $\alpha$ is the usual time-averaged ratio of
vertically-integrated $r$-$\phi$ stress to vertically-integrated pressure.   The
transition radius would then be found at $R_T \sim \Phi  a_*^{2/3} (h/R_T)^{-4/3} r_g$.

\section*{Acknowledgements}
We would like to thank Cole Miller and Steve Lubow for extensive and valuable discussions.
This work was partially supported under National Science Foundation grants AST-1028111
and AST-0908326 (JHK and KAS), AST-0908869 (JFH), and NASA grant NNX09AD14G (JFH).
The National Science Foundation also supported this research in part through XSEDE resources
on the Kraken cluster through Teragrid allocation TG-MCA95C003.

\bibliographystyle{apj}
\bibliography{Bib}

\section{Appendix}

\begin{figure}
\begin{center}
\includegraphics[width=0.6\textwidth,angle=90]{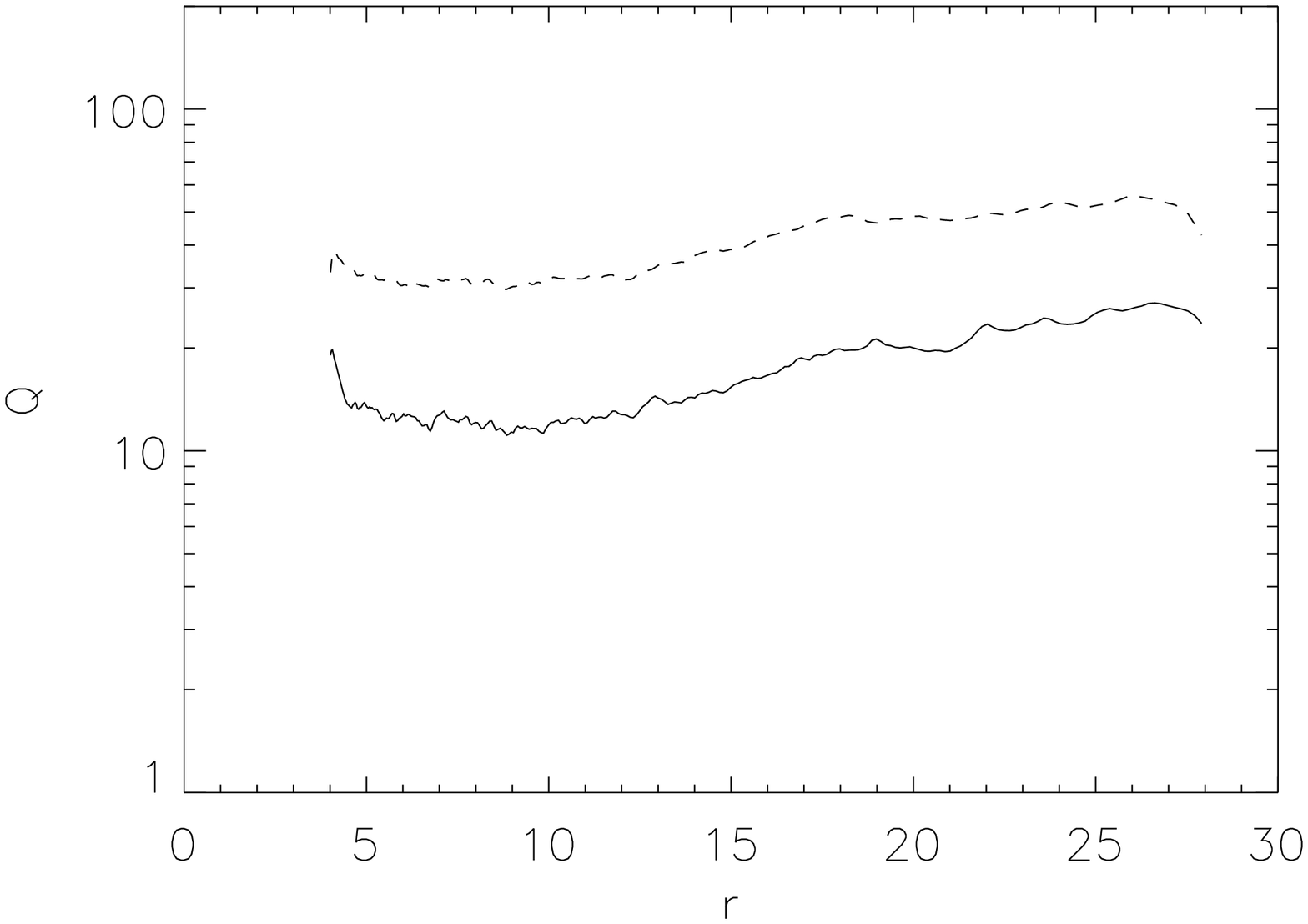} \\
\includegraphics[width=0.6\textwidth,angle=90]{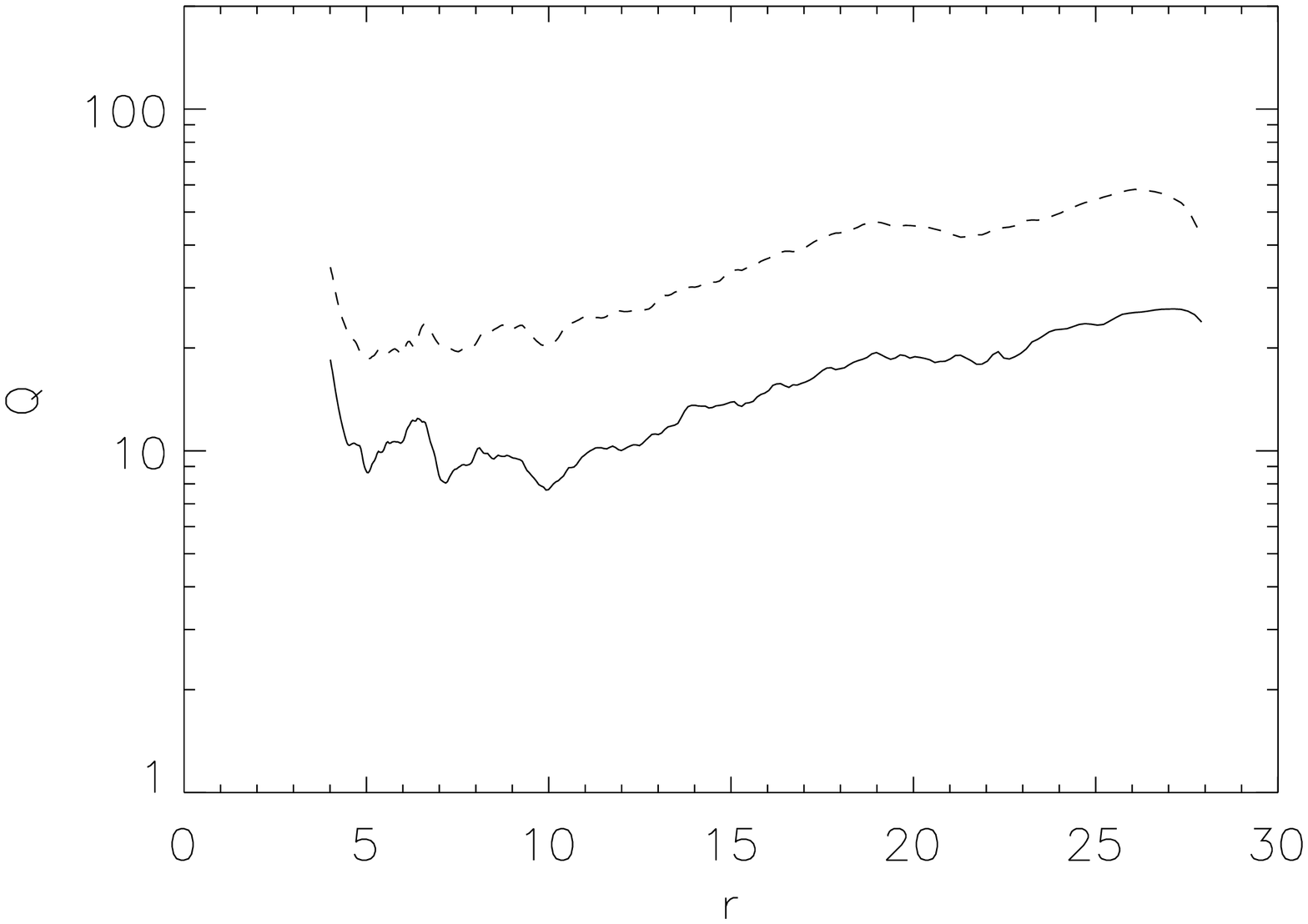} \\
\caption{Density-weighted $Q_z$ (solid curve) and $Q_\phi$ (dashed curve) at times
$t=14.5$ (upper panel) and $t=17$ (lower panel) in \runm.  We define $Q_\phi$ as
the quality factor in the azimuthal direction in the coordinate equatorial plane.}
\label{fig:quality}
\end{center}
\end{figure}

  In Figure~\ref{fig:quality}, we show the density-weighted MHD resolution
quality factors as functions of radius in \runm\ at two times, one during the saturated
MHD turbulence immediately prior to the beginning of the torques ($t=14.5$) and one well
into the evolution with torques ($t=17$).  The quality factors are defined as
\begin{equation}
Q_{x} = \frac{2\pi v_{A,x}}{\Delta x},
\end{equation}        
where $v_{A,x}$ is the Alfven speed restricted to the $x-$component of the magnetic field.
Although the disk tilts out of the equatorial plane of the coordinate system as
\runm\ evolves past $t=15$, at $t=17$, only the smallest radii, $r \lesssim 6$ have
changed their orientation in a noticeable way, so we make the approximation that the
directions of the axes for the disk remain the coordinate axes.

At $t=14.5$, $\langle Q_z\rangle_{\rho} \simeq 12$--25, while
$\langle Q_{\phi}\rangle_{\rho} \simeq 30$--50.  Although the magnetic field is
significantly weakened immediately after the torques begin, at $t=17$ the
quality factors are still fairly good: $\langle Q_z\rangle_{\rho} \simeq 8$--20, while
$\langle Q_{\phi}\rangle_{\rho} \simeq 20$--50.  We also checked later times and found
that they were not much different from $t=17$ in terms of these numbers.
\cite{HGK11,Sorathia12} recommended
that both $Q_z$ and $Q_\phi$ should be $\gtrsim 10$ and preferably $\gtrsim 20$,
but also remarked that a particularly large value of one could compensate for
a smaller value of the other.  On this basis, we regard the simulation as reasonably
well-resolved throughout.
However, it is possible, particularly when larger inclination angles are explored,
that the quality factors required for resolving MRI-driven MHD turbulence may be different
in warped disks.

\end{document}